\documentclass[amsmath,amssymb,showpacs,nofootinbib,11pt]{revtex4-1}
\usepackage{graphicx}
\usepackage{dcolumn}
\usepackage{bm}
\usepackage{color} 
\usepackage{slashed}
\usepackage{tabu}
\usepackage{float} 
\allowdisplaybreaks
\newcommand{\hs}{\hspace*{0.5cm}}
\newcommand{\vs}{\vspace*{0.5cm}}
\newcommand{\be}{\begin{equation}}
\newcommand{\ee}{\end{equation}}
\newcommand{\bea}{\begin{eqnarray}}
\newcommand{\eea}{\end{eqnarray}}
\newcommand{\ben}{\begin{enumerate}}
\newcommand{\een}{\end{enumerate}}
\newcommand{\bde}{\begin{widetext}}
\newcommand{\ede}{\end{widetext}}

\newcommand{\crn}{\nonumber \\}

\newcommand{\al}{\alpha}

\newcommand{\ga}{\gamma}

\newcommand{\om}{\omega}
\newcommand{\pa}{\partial}

\newcommand{\fr}{\frac}
\newcommand{\bc}{\begin{center}}
\newcommand{\ec}{\end{center}}
\newcommand{\Ga}{\Gamma}

\newcommand{\La}{\Lambda}

\newcommand{\Om}{\Omega}
 
\newcommand{\AdrHEPC}{Phenikaa Institute for Advanced Study and Faculty of Basic Science, Phenikaa University, Yen Nghia, Ha Dong, Hanoi 100000, Vietnam}
 
\begin{document} 

\title{Dequantization of electric charge: Probing scenarios of cosmological multi-component dark matter}
\author{Duong Van Loi}
\author{Nguyen Manh Duc\footnote{Leave of absence from Hanoi--Amsterdam High School for the Gifted, Trung Hoa, Cau Giay, Hanoi 100000, Vietnam}}
\author{Phung Van Dong}
\email{Corresponding author; email: dong.phungvan@phenikaa-uni.edu.vn}
\affiliation{\AdrHEPC}  

\date{\today}

\begin{abstract}
Since the electric charge in the standard model is theoretically not quantized, we may have a variant of it, called dark charge. Similar to the electric charge, the dark charge neither commutes nor closes algebraically with $SU(2)_L$. The condition of algebraic closure leads to a novel gauge extension, $SU(2)_L \otimes U(1)_Y \otimes U(1)_N$, where $Y$ and $N$ determine the electric and dark charges, respectively, apart from the color group. We argue that the existence of the dark charge, thus $N$, leads to novel scenarios of multi-component dark matter, in general. The dark matter stability is determined by a residual (or dark charge) gauge symmetry isomorphic to an even $Z_k$ discrete group, where $k$ is specified dependent on the value of the neutrino dark charge. This residual symmetry divides the standard model particles into distinct classes, which possibly accommodate dark matter, but each dark matter candidate cannot decay due to the color and electric charge conservation. We analyze in detail three specific models according to $k=2,4,6$ and determine the simplest dark matter candidates. For small $U(1)_N$ coupling, the two-component dark matter scenarios implied by the dark charge successfully explain the dark matter relic density and the recent XENON1T excess, as well as the beam dump, neutrino scattering, and astrophysical bounds. Otherwise, for large $U(1)_N$ coupling, we have multi-WIMPs coexisted beyond the weak scale.
\end{abstract}


\maketitle

\section{Introduction}

The standard model of particle physics \cite{Zyla:2020zbs} has been enormously successful in describing most of the observed phenomena. However, the explicit evidences of neutrino oscillation \cite{RevModPhys.88.030501,RevModPhys.88.030502} and dark matter existence \cite{Hinshaw:2012aka,Aghanim:2018eyx} cannot be explained within the framework of the standard model. This proves that the standard model must be extended in order to account for the new physics beyond.

The existing experiments of dark matter relic density, direct/indirect detections, and particle colliders have not revealed any detail of particle picture of dark matter. Obviously, many dedicated theories often assume dark matter to be composed of a single particle kind, such as a WIMP \cite{Jungman:1995df,Bertone:2004pz}, while there is no any reason why only a single particle kind of dark matter is presented. Since the constituent of dark matter is still eluded, a scheme of multi-component dark matter seems to be naturally in comparison to the rich structure of normal matter within an atom. Indeed, the possibility that dark matter consists of more than one type of particle is very attractive~\cite{Berezhiani:1989fp,Berezhiani:1990sy,Ma:2006uv,Zurek:2008qg,Fukuoka:2010kx,Batell:2010bp,Chialva:2012rq,Biswas:2013nn,Bhattacharya:2013hva,Bian:2013wna,Belanger:2014bga,Bian:2014cja,Esch:2014jpa,DiFranzo:2016uzc,DuttaBanik:2016jzv,Karam:2016rsz,Bhattacharya:2016ysw,Arcadi:2016kmk,Borah:2017xgm,Ahmed:2017dbb,Bhattacharya:2017fid,Bhattacharya:2018cgx,Bhattacharya:2018cqu,Aoki:2018gjf,DuttaBanik:2018emv,Barman:2018esi,YaserAyazi:2018lrv,Chakraborti:2018aae,Chakraborti:2018lso,Borah:2019epq,Elahi:2019jeo,Yaguna:2019cvp}, with intriguing results for galaxy structure \cite{Fan:2013yva,Fan:2013tia}, multiple gamma-ray line \cite{Boehm:2003ha}, boosted dark matter \cite{Agashe:2014yua,Kong:2014mia,Alhazmi:2016qcs,Kim:2016zjx,Giudice:2017zke,Chatterjee:2018mej,Kim:2018veo}, and dark matter self-interaction \cite{Elbert:2014bma,Tulin:2017ara}, as well as those motivated with neutrino mass generation~\cite{Heeck:2012bz,Aoki:2012ub,Aoki:2013gzs,Kajiyama:2013rla,Karam:2015jta,Bernal:2018aon,Bonilla:2018ynb,Borah:2019aeq,Bhattacharya:2019fgs,Biswas:2019ygr,Bhattacharya:2019tqq}.  

We wish to suggest that the structure of dark matter and the origin of neutrino mass can be simultaneously understood from a very fundamental aspect of the standard model, that is the electrodynamics. As shown in \cite{VanDong:2020cjf}, since the electric charge ($Q$) is theoretically not fixed, there may exist a variant of it, interpreted to be a physical dark charge ($D$), which coexists with the electric charge, driving the model extension with implication for neutrino mass and dark matter. As an indication, the most $U(1)$ gauge extensions that mix with the hypercharge ($Y$) actually imply such a dark charge to be a dequantization version of the usual electric charge. Analogous to the electric charge, the dark charge neither commutes nor closes algebraically with the $SU(2)_L$ weak isospin, $T_i$ ($i=1,2,3$). By algebraic closure condition, or symmetry (flipping) principle, this leads to the full gauge symmetry, $SU(3)_C\otimes SU(2)_L \otimes U(1)_Y \otimes U(1)_{N}$, where the last two $U(1)$'s factors determine the electric charge and the dark charge through the $T_3$ operator, i.e. $Y=Q-T_3$ and $N=D-T_3$, respectively, where the QCD group has been included for completeness. Additionally, to cancel the $U(1)_N$ anomalies, three right-handed neutrino singlets must be included, which have a dark charge identical to that of the usual neutrinos, whereas both kinds of the neutrinos have vanished electric charge \cite{VanLoi:2020kdk}. These right-handed neutrino singlets obtain large Majorana masses according to the $N$-charge breaking, while they couple to the usual neutrinos via the Higgs field. As a result, appropriate small neutrino masses are induced in terms of a canonical seesaw mechanism. Especially, the seesaw scale that breaks the $N$-charge, combined with the weak scale which breaks the weak isospin $T_i$, defines a residual discrete symmetry of the dark charge $D=T_3+N$, which manifestly stabilizes dark matter candidates.\footnote{Dark matter stability mechanism is different from the matter parity that results as a residual gauge symmetry, see e.g. \cite{Dong:2013wca,Dong:2015yra,Alves:2016fqe,Dong:2016gxl,Dong:2017zxo,Dong:2018aak,Nam:2020twn,VanDong:2020bkg,VanLoi:2019eax,VanLoi:2020xcq}, because the dark charge does not commute with the weak isospin, while the matter parity induced by $B-L$ breaking does.}

In the literature, to make dark matter candidates co-existed (i.e. co-stabilized) responsible for multi-component dark matter, discrete groups that are larger than a parity, e.g. a non-simple $Z_2\otimes Z'_2$ group, have often been assumed by {\it ad hoc}. We show in this work that such stability mechanism originates from the residual gauge symmetry of the dark charge, as mentioned, and that it generically leads to novel scenarios of multi-component dark matter. Indeed, depending on the value of the neutrino dark charge, $\delta$, the residual gauge symmetry may be correspondingly isomorphic to a discrete group $Z_k$, where $k$ is an even integer. Such a residual symmetry divides the standard model particles into various classes, which may accommodate dark matter candidates to be simultaneously stabilized. We analyze in detail three concrete models according to $k=2,4,6$. The model with $k=2$ implies a single dark matter candidate, while the remaining two cases for $k=4,6$, as well as the models with a larger value of $k$, provide multi-component dark matter. We prove that the dark matter candidates can have an arbitrary mass---not necessarily lighter than the charged leptons and quarks---although both the dark matter and the normal matter transform nontrivially under the same residual symmetry, a consequence of the electric and color charge conservation. This is opposite to the previous announcement \cite{VanDong:2020cjf} as well as differing from the most extensions for dark matter. However, the dark matter candidates in the model with $k=4$, except for the case of scalar dark matter, must be lighter than the $W$ boson, due to the dark field self-interactions that potentially lead to a dark matter instability. As multi-component dark matter is commonly implied, we choose the models with $k=4,6$ for further experimental investigation. We figure out the viable parameter regime satisfying the dark matter abundance, direct detections, particle colliders, as well as astrophysical bounds.

The rest of this work is organized as follows. In Sec. \ref{model}, we construct the model for a generic dark charge. In Sec. \ref{parity}, we determine the residual discrete symmetry, showing that it may be isomorphic to a $Z_k$ with even $k$ value. Then, the simplest candidates of multi-component dark matter are identified according to several $k$ values. In Sec. \ref{pheno2c}, we present the phenomenology of the model, including neutrino mass, $Z'$ constraint, $U(1)_N$ running coupling, and that the schemes with $k=4,6$ responsible for two-component dark matter are experimentally investigated. Finally, we summarize the results in Sec.~\ref{conl}.

\section{\label{model} A model of dark charge}

In the nature, electric charges of fundamental particles come in discrete amounts, equal to integer multiples of an elementary electric charge, called electric charge quantization. However, if the electric charge is theoretically not quantized, i.e. not discrete, it is dequantized, i.e. arbitrary or continuous. This term is eventually associated with the theories, such as the electrodynamics and the standard model, that do not explain the quantization of electric charge. The former theory obeys the charge quantization, if a magnetic monopole is presented, proposed by Dirac long ago~\cite{Dirac:1931kp}. But, the monopole has not been observed yet. The latter theory implies the charge quantization, if all hidden symmetries, e.g. $L_a -L_b$ for $a,b=e,\mu,\tau$ and $B-L$, are violated. Otherwise, they make the hypercharge, thus the electric charge, free, such that $Y\rightarrow Y+x_{a b}(L_a-L_b)+y(B-L)$ is always allowed, called dequantization effect \cite{Babu:1989ex,Foot:1990uf}. 

Indeed, in the standard model with the gauge symmetry, \be SU(3)_C\otimes SU(2)_L\otimes U(1)_Y,\ee the electric charge is related to the hypercharge through the weak isospin by  
\be Q=T_3+Y.\label{add1dt}\ee The standard model does not predict the quantization of the electric charge because---while the value of $T_3$ is quantized due to the non-Abelian nature of $SU(2)_L$ Lie algebra---the value of $Y$ is completely arbitrary on the theoretical ground of $U(1)_Y$ group, since $[Y,Y]=0$ for any $Y$. For phenomenological purpose, $Y$ is often chosen to describe the observed electric charge values, while does not explain them. Additionally in the theoretical side, the value of $Y$ can be constrained by the anomaly cancellation and the fermion mass generation for each family, but it is dequantized when including three families or right-handed neutrinos into account \cite{Pisano:1996ht,Doff:1998we,deSousaPires:1998jc,deSousaPires:1999ca,VanDong:2005ux}. The reason is that the standard model always contains anomaly-free hidden symmetries, such as $L_a-L_b$ for $a,b=e,\mu,\tau$ or $B-L$ for including the right-handed neutrinos, that subsequently make $Y$ free, as mentioned. 

We consider the latter by introducing the three right-handed neutrinos, $\nu_{aR}$, to the standard model and assume $Y(\nu_{aR})=\delta$. The conditions of the anomaly cancellation and the fermion mass generation require $Y(e_{aR})=\delta-1$, $Y(l_{aL})=\delta-1/2$, $Y(u_{aR})=(2-\delta)/3$, $Y(d_{aR})=-(1+\delta)/3$, and $Y(q_{aL})=(1-2\delta)/6$, where $l_{aL}\equiv (\nu_{aL}\ e_{aL})^T$ and $q_{aL}\equiv (u_{aL}\ d_{aL})^T$ \cite{VanDong:2020cjf,VanLoi:2020kdk}.\footnote{We redefine $a$ (and $b$) to be a family index, that labels both leptons and quarks, without confusion.} We call $\delta$ to be the parameter of dequantization. Namely, for $\delta=0$, the hypercharge and the electric charge are properly recovered as in the normal sense, so we relabel $Y=Y|_{\delta=0}$ and $Q=Q|_{\delta=0}$ for convenience in reading. Whereas, for $\delta\neq 0$, we obtain the new charges as corresponding variants of the electric charge and the hypercharge, by which we define $D=Q|_{\delta\neq 0}$, called dark charge, and $N=Y|_{\delta\neq 0}$, called hyperdark charge. One has a similar relation to Eq.~(\ref{add1dt}), \be D=T_3+N.\label{charges}\ee 
As the electric charge, the dark charge $D$ neither commutes nor closes algebraically with the weak isospin, which indicates to a novel gauge extension, such as
\be
SU(3)_C\otimes SU(2)_L\otimes U(1)_Y\otimes U(1)_N \label{gausym},
\ee where $N$ thus determines $D$ in the same way that $Y$ does so for $Q$, respectively. 

It is easily realized that $N$ was studied in the literature as a combination of $x Y +y (B-L)$ for $x=1$ and $y=- \delta$, since both $Y,B-L$ are anomaly-free. However, the interpretation of the dark charge $D$ as well as this combination to be a dequantization of the electric charge did not appear until our proposal \cite{VanDong:2020cjf,VanLoi:2020kdk}. An important result is that $D$ implies dark matter stability; additionally, the dark matter can be unified with normal matter in $SU(2)_L$ multiplet, because the dark charge $D$ is noncommutative, which differs from the most extensions. Let us stress that dark matter must be stabilized over the cosmological timescales. For instance, if a dark matter particle with a mass between the weak and TeV scales decays to two lighter particles, the relevant coupling strength that couples these fields must be around $\sim 10^{-20}$. Indeed, the current bound of dark matter lifetime is at least a billion times longer than our universe's age, leading to a corresponding coupling strength $\sim 10^{-25}$ \cite{Fermi-LAT:2012ugx}. Such tiny coupling, that is much smaller than the Dirac neutrino coupling $\sim 10^{-12}$, is not naturally stabilized against quantum contributions, which might only be understood by symmetries.   

The full fermion content supplied with the hyperdark charge under the gauge symmetry in Eq. (\ref{gausym}) is 
\bea l_{aL} &=& \left(\begin{array}{c}\nu_{aL}\\ e_{aL}\end{array}\right)\sim \left(1,2,-\fr 1 2, \delta-\fr 1 2\right),\\
q_{aL} &=& \left(\begin{array}{c}u_{aL}\\ d_{aL}\end{array}\right)\sim \left(3,2,\fr 1 6, \fr 1 6-\fr{\delta}{3}\right),\\
\nu_{aR}&\sim & (1,1,0,\delta),\\
e_{aR}&\sim & (1,1,-1,\delta-1),\\
u_{aR} &\sim &(3,1, 2/3,2/3-\delta/3),\\
d_{aR}&\sim &(3,1,-1/3,-\delta/3-1/3),\eea where $a=1,2,3$ is a family index, and $\delta$ is arbitrarily nonzero, $\delta\neq 0$. The left and right fermion $f_{L,R}$ have the same dark charge, which coincides with the hyperdark charge value of the right component above, since $D(f)=D(f_R)=N(f_R)$ due to $T_3(f_R)=0$. 

It is easily verified that all the anomalies are cancelled out for each family, independent of $\delta$, as shown in Appendix of \cite{VanLoi:2020kdk}. There, we also indicated that the anomalies are cancelled out for different dequantizations, such as $U(1)_N\rightarrow U(1)_{N_1}\otimes U(1)_{N_2}\otimes \cdots \otimes U(1)_{N_p}$, according to $\delta\rightarrow \delta_1,\delta_2,\cdots,\delta_p$, respectively. Multi-component dark matter is naturally recognized, since each $U(1)_{N}$'s factor presents an independent stability mechanism (see below). However, within one factor, the $U(1)_N$ breaking also supplies the schemes of multi-component dark matter. That said, the multi-component dark matter is a generic result of the electric charge dequantization. Hereafter, only the latter case is taken into account.         

To break the gauge symmetry and generate appropriate masses, the scalar content is introduced as 
\be 
\phi=\left(\begin{array}{c}\phi^+\\ \phi^0\end{array}\right)\sim \left(1,2,\fr 1 2, \fr 1 2\right), \hs \chi\sim (1,1,0,-2\delta).
\ee Here, the new field $\chi$ is necessarily to break $U(1)_N$, producing Majorana $\nu_R$ masses via the couplings $\chi \nu_R\nu_R$. Additionally, $\phi$ is the usual Higgs doublet, which has the dark charge identical to the electric charge, such that both of these charges are not broken by the weak vacuum. The vacuum expectation values (VEVs) are given by \be\langle\phi\rangle= \left(\begin{array}{c}
0\\ \fr{v}{\sqrt{2}}\end{array}\right),\hs \langle\chi\rangle=\fr{\Lambda}{\sqrt{2}},\ee satisfying \be\Lambda\gg v=246 \text{ GeV},\ee for consistency with the standard model.

\section{\label{parity} Residual dark charge}
The gauge symmetry is broken via two stages,
\bc
\begin{tabular}{c}
$SU(3)_C\otimes SU(2)_L\otimes U(1)_Y\otimes U(1)_N$\\
$\downarrow \La $\\
$SU(3)_C\otimes SU(2)_L\otimes U(1)_Y\otimes R_N$\\
$\downarrow v$\\
$SU(3)_C\otimes U(1)_Q\otimes R_D.$
\end{tabular}
\ec

In the first stage, only $U(1)_N$ is broken by the VEV of $\chi$, because of $N \langle\chi\rangle=-\sqrt{2}\delta\Lambda\neq 0$. But, this breaking is not completed. Indeed, a residual symmetry of $N$ is defined as $R_N=e^{i\alpha N}$, since it is a $U(1)_N$ transformation. $R_N$ conserves the vacuum, $R_N\langle\chi\rangle=\langle\chi\rangle$, leading to $e^{-i2\alpha\delta}=1$, or $\alpha=k\pi/\delta$ for $k$ integer. We deduce \be R_N=e^{ik\pi N/\delta}=(-1)^{kN/\delta}.\label{gadttn1}\ee

In the second stage, the VEV of $\phi$ breaks all $SU(2)_L\otimes U(1)_Y\otimes R_N$, except for the color group, because of $T_i\langle\phi\rangle=\fr 1 2\sigma_i\langle\phi\rangle\neq 0$ and $Y\langle\phi\rangle=N\langle\phi\rangle=(0\ v/2\sqrt{2})^T\neq 0$. However, the breaking is also not completed. Indeed, a conserved charge after the breaking must take the form, \be X\equiv a_i T_i+bY+cN,\ee that annihilates the weak vacuum, $X\langle\phi\rangle=0$. This yields $a_1=a_2=0$ and $a_3=b+c$, hence \bea X &=& b(T_3+Y)+c(T_3+N)\crn
&=& b Q+cD,\eea where $Q=T_3+Y$ and $D=T_3+N$ are defined as before. Additionally, $Q,D$ commute, $[Q,D]=0$, and separately conserve the weak vacuum (similar to $X$), i.e. $Q\langle\phi \rangle=D\langle\phi\rangle=0$. Hence, the residual symmetry $X$ is Abelian by itself and factorized into \be U(1)_X=U(1)_Q\otimes U(1)_D,\ee according to a transformation, $e^{iX}=e^{i b Q} e^{i c D}$, multiplied by those of $Q,D$, respectively.

The first factor $U(1)_Q$ is well known to be the electromagnetic symmetry, which is the residual symmetry of $SU(2)_L\otimes U(1)_Y$, obviously conserving the $\chi$ vacuum, since $Q\langle \chi\rangle =0$. The second factor $U(1)_D$ is a residual symmetry of $SU(2)_L\otimes R_N$, which similar to $R_N$, must conserve the $\chi$ vacuum. In other words, $e^{i c D}\langle\chi\rangle=\langle\chi\rangle$ leads to $e^{-i2c \delta}=1$, implying $c=\alpha =k\pi/\delta$. This restricts $U(1)_D$ to $R_D$ defined by 
\be
R_D=e^{ik\pi D/\delta}=(-1)^{kD/\delta},\label{gadttn}
\ee to be a residual symmetry of the dark charge $D=T_3+N$, thus of $SU(2)_L\otimes U(1)_N$. 

It is clear that $R_D$ is shifted from $R_N$ by the weak vacuum, not commuted with the standard model symmetry, a consequence of the existence of noncommutative dark charge. $R_D$ would distinguish particles with different weak isospin values and give a potential unification of normal matter and dark matter in the same gauge multiplet \cite{VanDong:2020cjf}. This feature compares with supersymmetry, which does so for both kinds of matter (with different spin values) in supermultiplet. $R_D$ differs from the most $U(1)$ extensions in the literature, which actually lead to commutative discrete symmetries, like $R_N$, by contrast.   

For comparison we collect the electric and dark charges as well as the $R_D$ values of all fields in Table \ref{QDRD}, where $A$ commonly denotes all the gauge fields corresponding to the gauge symmetry in Eq. (\ref{gausym}), except for the $W$ boson. Additionally, the generation and left/right chirality indices have been omitted since the relevant fields have the same $Q$, $D$, and $R_D$ values, respectively.
\begin{table}[h]
\bc
\caption{$Q$, $D$, and $R_D$ values of fields.}
\label{QDRD}  
\begin{tabular}{llll}
\hline\noalign{\smallskip}
Field & $Q$ & $D$ & $R_D$ \\
\noalign{\smallskip}\hline\noalign{\smallskip}
$\nu$ & $0$ & $\delta$ & $(-1)^k$ \\
$e$ & $-1$ & $\delta-1$ & $(-1)^{k(\delta-1)/\delta}$ \\
$u$ & $2/3$ & $(2-\delta)/3$ & $(-1)^{k(2-\delta)/3\delta}$ \\
$d$ & $-1/3$ & $-(1+\delta)/3$  & $(-1)^{-k(\delta+1)/3\delta}$ \\
$\chi$ & $0$ & $-2\delta$ & $1$ \\
$\phi^+,W^+$ & $1$ & $1$ & $(-1)^{k/\delta}$ \\
$\phi^0,A$ & $0$ & $0$ & $1$\\ 
\noalign{\smallskip}\hline
\end{tabular}
\ec
\end{table}

From Table \ref{QDRD}, it is clear that if $k=0$, then $R_D=1$ for all fields and every $\delta$, which is the identity transformation. To search for the residual group structure, we find the minimal value of $|k| \neq 0$ that still satisfies $R_D=1$ for all fields. Such value of $k$ depends on $\delta$ and obeys the following equations simultaneously,
\bea
k&=&2n_1,\\
k(\delta-1)/\delta &=&2n_2,\\
k(\delta-2)/3\delta &=&2n_3,\\
k(\delta+1)/3\delta &=&2n_4,\\
k/\delta &=&2n_5,
\eea
where $n_{1,2,3,4,5}=0,\pm 1,\pm 2,\cdots$ are integer. From the first equation, $k$ must be even integer, i.e. $k=\pm 2, \pm 4, \pm 6, \cdots$. 

First, considering the case of $|k| = 2$, we get $n_1 = \pm 1$ and subsequent relations
\be
n_2=2-3n_4,\hs n_3=1-2n_4,\hs n_5=3n_4-1.
\ee
We also obtain a corresponding value of $\delta$, denoted by
\be
\delta_2 = \frac{1}{3n_4-1} =-1,\frac{1}{2},-\frac{1}{4},\frac{1}{5},-\frac{1}{7},\frac{1}{8},-\frac{1}{10},\frac{1}{11},\cdots.
\ee

Similarly, considering the next cases with $|k| = 4$ and 6, we obtain the corresponding values of $\delta$, denoted by 
\bea
\delta_4 &=& \frac{2}{3n_4-2}=-1,2,-\frac{2}{5},\frac{1}{2},-\frac{1}{4},\frac{2}{7},\frac{1}{5},-\frac{2}{11},\cdots,\\
\delta_6 &=& \frac{1}{n_4-1}=\pm 1,\pm \fr{1}{2},\pm \fr{1}{3},\pm \fr{1}{4},\pm \fr{1}{5},\pm \fr{1}{6},\pm \fr{1}{7},\cdots,
\eea respectively. Notice that $\delta_4$ contains in part the values of $\delta_2$ which correspondingly reduce to the first case and must be omitted. Moreover, $\delta_6$ also includes values of $\delta_2$ and a part of $\delta_4$ which reduce to the first or the second case and must be eliminated too. 

In other words, there are different values of $\delta$ resulting in $R_D = 1$ for all fields according to the minimal solution of $|k|=2$, $4$, or $6$, except the identity $k=0$. This implies that the residual symmetry $R_D$ is automorphic to a discrete group, $Z_2$, $Z_4$, or $Z_6$, respectively, providing dark matter candidates. To be concrete, in Table \ref{5RD}, we show three specific cases, giving the results as expected. Notice that unlike the minimal value of $k$ according to $R_D=1$ for all fields, we have restored $k$, without confusion, to be arbitrarily integer characterizing transformations of (\ref{gadttn}) within each residual group $R_D$. 
\begin{table}[h]
\bc
\caption{$R_D$ values of all fields when $\delta$ is fixed, where $k$ is arbitrarily integer.}
\label{5RD}  
\begin{tabular}{llll}
\hline\noalign{\smallskip}
Field & $R_D\rightarrow Z_2$ & $R_D\rightarrow Z_4$ & $R_D\rightarrow Z_6$ \\
& $(\delta=-1)$ & $(\delta=2)$ & $(\delta=1)$ \\
\noalign{\smallskip}\hline\noalign{\smallskip}
$\nu$ & $(-1)^k$ & $(-1)^k$ & $(-1)^k$ \\
$e$ & $1$ & $(-1)^{k/2}$ & $1$ \\
$u$ & $(-1)^{-k}$ & $1$ & $(-1)^{k/3}$ \\
$d$ & $1$ & $(-1)^{-k/2}$ & $(-1)^{-2k/3}$ \\
$\chi$ & $1$ & $1$ & $1$ \\
$\phi^+,W^+$ & $(-1)^{-k}$ & $(-1)^{k/2}$ & $(-1)^k$ \\
$\phi^0,A$ & $1$ & $1$ & $1$ \\ 
\noalign{\smallskip}\hline
\end{tabular}
\ec
\end{table}

Of course, there are values of $\delta$ that lead to the residual symmetries higher than the given ones, such as $Z_8$, $Z_{10}$, $Z_{12}$, and so forth. They would imply more the stabilized dark matter components than the chosen ones. Such generalization to a higher discrete symmetry is not trivial, but not discussed further in this work.
  
\subsection{The model with $\delta=-1$}

The second column of Table \ref{5RD} indicates $R_D=1$ for all fields and for the minimal value of $|k|=2$, except $k=0$. Hence, the residual symmetry $R_D$ is automorphic to 
\be Z_2=\lbrace 1,g\rbrace,\ee 
where $g\equiv e^{-i\pi D} =(-1)^{-D}$ and $g^2=1$. 

Because the spin parity, $P_S=(-1)^{2s}$, is always conserved by the Lorentz symmetry, we conveniently multiply the residual symmetry with the spin parity group $S=\lbrace1,P_S\rbrace$ with $P_S^2=1$, to perform a new group $Z_2\otimes S$, which has an element
\be g_S=g\times P_S=(-1)^{-D+2s}, \ee
satisfying $g_S^2=1$. Therefore, we have
\be \mathcal{Z}_2=\lbrace 1,g_S\rbrace \ee
to be an invariant subgroup of $Z_2\otimes S$, and decompose
\be Z_2\otimes S\cong[(Z_2\otimes S)/\mathcal{Z}_2]\otimes \mathcal{Z}_2. \ee
Since $[(Z_2\otimes S)/\mathcal{Z}_2]= \lbrace\lbrace 1,g_S \rbrace,\lbrace g,P_S \rbrace\rbrace$ is conserved if $g_S$ is conserved, we can omit it, interpreting the $\mathcal{Z}_2$ group as a residual symmetry instead of $Z_2$. 

The field representations under $\mathcal{Z}_2$ are summarized in Table \ref{Ggroup}. Notice that $\mathcal{Z}_2$ has two one-dimensional irreducible representations, $\underline{1}^{(1)}$ and $\underline{1}^{(2)}$, corresponding to $g_S=1$ and $-1$, respectively, as usual. 
\begin{table}[h]
\bc
\caption{Field representations under the $\mathcal{Z}_2$ group.}
\label{Ggroup}  
\begin{tabular}{lrl}
\hline\noalign{\smallskip}
Field & $g_S$ & $\mathcal{Z}_2$\\
\noalign{\smallskip}\hline\noalign{\smallskip}
$\nu$ & $1$ & $\underline{1}^{(1)}$\\
$e$ & $-1$ & $\underline{1}^{(2)}$\\
$u$ & $1$ & $\underline{1}^{(1)}$\\
$d$ & $-1$ & $\underline{1}^{(2)}$\\
$\chi$ & $1$ & $\underline{1}^{(1)}$\\
$\phi^+,W^+$ & $-1$ & $\underline{1}^{(2)}$\\
$\phi^0,A$ & $1$ & $\underline{1}^{(1)}$\\ 
\noalign{\smallskip}\hline
\end{tabular}
\ec
\end{table}

From Table \ref{Ggroup}, the model with $\delta=-1$ implies scenarios of a single dark field, labeled $\Psi_1$, which transforms nontrivially under  $\mathcal{Z}_2$, i.e. $g_S=-1$, such as
\be \Psi_1\sim (1,1,0,2d_1)\ee
for a fermion or
\be \Psi_1\sim (1,1,0,2d_1+1)\ee
for a scalar, where $d_1$ is integer. For simplicity, throughout this work, we assume dark matter candidates to be spin-0 bosonic and/or spin-$1/2$ fermionic fields, transforming as the standard model singlet.\footnote{Introduction of $SU(2)_L$ multiplets leads to a unification of normal field and dark field by the gauge symmetry due to the noncommutative $R_D$, as mentioned. This possibility is interesting, but is out of the scope of this work and skipped.} This result is similar to the previous interpretation obtained in \cite{VanDong:2020cjf,VanLoi:2020kdk}. Therefore, we obtain the simplest dark matter candidate to be either a fermion or a scalar with $d_1=0$.

We would like to stress that $\Psi_1$ can have an arbitrary mass, which does not decay to the usual particles. First, all $\Psi_1$, $e$, $d$, $\phi^+$, and $W^+$ are $\mathcal{Z}_2$ odd. Next, $\Psi_1$ is color and electrically neutral, whereas the rest ($e, d, \phi^+, W^+$) carry a color and/or an electric charge. The color and electric charge conservation demands that $\Psi_1$ cannot decay to such a colored and/or electrically charged state. Indeed, by contrast, suppose that $\Psi_1$ decays to a final state. This state must be color and electrically neutral, as $\Psi_1$ is. Additionally, the $\mathcal{Z}_2$ conservation requires that the final state must be combined of an odd number of the odd fields $(e,d,\phi^+, W^+)$. Since $\phi^+$ is eaten by $W^+$ and that $W^+$ decays to $(e^+,\nu)$ or $(d^c,u)$, the final state includes only $(e,d)$ as possible old fields. Thus, the decay process takes the form, \be \Psi_1\rightarrow x e^-+\bar{x} e^+ + y d + \bar{y} d^c + z u + \bar{z} u^c+\cdots,\ee where the dots stand for the remaining fields (if any) that are electrically and color neutral and $\mathcal{Z}_2$ even. The conservation laws imply, i) $x+\bar{x}+y+\bar{y}=2n+1$ ($\mathcal{Z}_2$ odd), ii) $-x+\bar{x}-y/3+\bar{y}/3+2z/3-2\bar{z}/3=0$ (electrically neutral), and iii) $y+z-\bar{y}-\bar{z}=3 n'$ (color neutral), for $n,n'$ integer. The last two equations lead to $-x+\bar{x}-y+\bar{y}+2n'=0$, which combined with the first equation, yields a contradiction to be $2(\bar{x}+\bar{y}+n')=2n+1$. In other words, the symmetries $SU(3)_C$, $U(1)_Q$, and $\mathcal{Z}_2$ jointly suppress the decay of the $\Psi_1$ dark matter, if it has a mass larger than the ordinary odd particles ($e,d,\phi^+$, $W^+$). Hence, the $\Psi_1$ dark matter can have an arbitrary mass, and its stability is different from the most extensions.    

\subsection{The model with $\delta=2$}

From the third column of Table \ref{5RD}, it is clear that $R_D=1$ for all fields with minimal value of $|k|=4$, except $k=0$. Thus, the residual symmetry $R_D$ is automorphic to  
\be Z_4=\lbrace 1,t,t^2,t^3\rbrace,\ee 
where $t\equiv e^{i\pi D/2}=(-1)^{D/2}$ and $t^4=1$. 

For convenience, we multiply the residual symmetry with the spin-parity group $S=\{1,P_S\}$ to perform $Z_4\otimes S$, which has an invariant subgroup,
\be
\mathcal{Z}_4=\lbrace 1, t_S,t_S^2,t_S^3\rbrace,
\ee
where $t_S=t\times P_S=e^{i\pi (D/2+2s)}=(-1)^{D/2+2s}$ and $t_S^4=1$. Therefore, we decompose
\be
Z_4\otimes S\cong[(Z_4\otimes S)/\mathcal{Z}_4]\otimes \mathcal{Z}_4.
\ee
Since $[(Z_4\otimes S)/\mathcal{Z}_4]=\lbrace \lbrace 1, t_S,t_S^2,t_S^3\rbrace,\lbrace t,P_S,t^2\times P_S,t^3\rbrace\rbrace$ is conserved if $\mathcal{Z}_4$ is conserved, we can consider $\mathcal{Z}_4$ to be a residual symmetry instead of $Z_4$ (note that $P_S$ is always preserved). The field representations under $\mathcal{Z}_4$ are given in Table \ref{Z4group} (third column). Here, $\mathcal{Z}_4$ has four one-dimensional irreducible representations, $\underline{1}^{(1)}$ according to $t_S=1$, $\underline{1}^{(2)}$ according to $t_S=i$, $\underline{1}^{(3)}$ according to $t_S=-1$, and $\underline{1}^{(4)}$ according to $t_S=-i$, with $i$ to be imaginary unit.  
\begin{table}[h]
\bc
\caption{Field representations under $\mathcal{Z}_4$ and their transformations under $Z_2\otimes Z'_2$.}
\label{Z4group}  
\begin{tabular}{lccccc}
\hline\noalign{\smallskip}
Field & $t_S$ & $\mathcal{Z}_4$ & $t^2_S$ & $[t_S]$ & $Z_2\otimes Z'_2$\\
\noalign{\smallskip}\hline\noalign{\smallskip}
$\nu$ & $1$ & $\underline{1}^{(1)}$ & 1 & $\{1,1\}$ & $\underline{1}^{(1)}\otimes \underline{1}^{(1)}$ \\
$e$ & $-i$ &  $\underline{1}^{(4)}$ & $-1$ & $\{-i,i\}$ & $\underline{1}^{(2)}\otimes \underline{1}^{(2)}$ \\
$u$ & $-1$ & $\underline{1}^{(3)}$ & 1 & $\{-1,-1\}$ & $\underline{1}^{(1)}\otimes \underline{1}^{(1)}$ \\
$d$ & $i$ & $\underline{1}^{(2)}$ & $-1$ & $\{i,-i\}$ & $\underline{1}^{(2)}\otimes \underline{1}^{(2)}$ \\
$\chi$ & $1$ & $\underline{1}^{(1)}$ & 1 & $\{1,1\}$ & $\underline{1}^{(1)}\otimes \underline{1}^{(1)}$ \\
$\phi^+,W^+$ & $i$ & $\underline{1}^{(2)}$ & $-1$ &  $\{i,-i\}$ & $\underline{1}^{(2)}\otimes \underline{1}^{(2)}$  \\
$\phi^0,A$ & $1$ & $\underline{1}^{(1)}$ & 1 & $\{1,1\}$ & $\underline{1}^{(1)}\otimes \underline{1}^{(1)}$ \\ 
\noalign{\smallskip}\hline
\end{tabular}
\ec
\end{table}

It is noted that $\mathcal{Z}_4$ contains an invariant subgroup $Z_2=\{1,t^2_S\}$ with the corresponding quotient group $Z'_2=\mathcal{Z}_4/Z_2=\{\{1,t^2_S\},\{t_S,t^3_S\}\}\equiv \{[1],[t_S]\}$, where each coset element $[x]$ consists of two elements of $\mathcal{Z}_4$, the characteristic $x$ and the other $t^2_S x$.\footnote{Hereafter, because the models with $\delta=-1,2,1$ are distinct, the notations used for labeling discrete groups and their presentations may be identical among the models, without confusion.} However, $\mathcal{Z}_4$ is not isomorphic to $Z_2\otimes Z'_2$. Additionally, one can verify that the fields transform the same under $Z_2$ and $Z'_2$, identical to the previous model with $\delta=-1$, because $t^2_S$ is identical to $g_S$, while $[t_S]$ is always isomorphic to $t^2_S$. They are also included in Table \ref{Z4group} as the last three columns, for clarity.\footnote{Hereafter, although the representations denoted by corresponding dimensions, as underlined, may appear the same across different groups, they must be distinguished as always supplied under their own group, which should not be confused.} That said, the $\mathcal{Z}_4$ basis is necessarily to supply multi-component dark matter, while the $Z_2\otimes Z'_2$ group is basically reducible to the single dark matter. Extra comment is that if one allows two factors $U(1)_{N_1}\otimes U(1)_{N_2}$ according to $\delta\rightarrow \delta_{1,2}$, as mentioned, they yield a residual Klein group $Z^{(1)}_2\otimes Z^{(2)}_2$ with the factors determined by independent dark charges $D_{1,2}$, respectively. This supplies a scheme of two-component dark matter whose candidates transform nontrivially under the $Z_2$'s factors, respectively, similar to the present $Z_4$ with only one $\delta$.   

\begin{table}[h]
\bc
\caption{Three kinds of dark field implied by $\mathcal{Z}_4$.}
\label{darkfieldZ4}  
\begin{tabular}{lcccc}
\hline\noalign{\smallskip}
Field & $\mathcal{Z}_4$ & Fermion & Scalar\\
\noalign{\smallskip}\hline\noalign{\smallskip}
$\Psi_2$ & $\underline{1}^{(2)}$ & $(1,1,0,4d_2-1)$ & $(1,1,0,4d_2+1)$\\
$\Psi_3$ & $\underline{1}^{(3)}$ & $(1,1,0,4d_3)$ & $(1,1,0,4d_3+2)$\\
$\Psi_4$ & $\underline{1}^{(4)}$ & $(1,1,0,4d_4+1)$ & $(1,1,0,4d_4+3)$\\
\noalign{\smallskip}\hline
\end{tabular}
\ec
\end{table}
From Table \ref{Z4group}, the model with $\delta=2$ implies three kinds of dark field, as determined in Table \ref{darkfieldZ4}, where $d_{2,3,4}$ are integer. It is able to show that $\mathcal{Z}_4$ is the smallest (i.e. $Z_k$) cyclic symmetry that allows a scenario of two-component dark matter \cite{Yaguna:2019cvp,Belanger:2014bga,Batell:2010bp}. Among the solutions, we consider here the two scalar dark fields to be the simplest candidates of dark matter, as resupplied in Table~\ref{canZ4}.
\begin{table}[h]
\bc
\caption{The simplest dark matter candidates implied by $\mathcal{Z}_4$.}
\label{canZ4}  
\begin{tabular}{lccc}
\hline\noalign{\smallskip}
Field & Spin & $t_S$ & $\mathcal{Z}_4$ \\
\noalign{\smallskip}\hline\noalign{\smallskip}
$\Phi_2\sim (1,1,0,1)$ & $0$ & $i$ & $\underline{1}^{(2)}$ \\
$\Phi_3\sim (1,1,0,2)$ & $0$ & $-1$ & $\underline{1}^{(3)}$ \\
\noalign{\smallskip}\hline
\end{tabular}
\ec
\end{table}

In this case of the model, the usual fields $e$, $u$, $d$, $\phi^+$, and $W^+$ transform nontrivially under $\mathcal{Z}_4$, analogous to the dark fields. Since $\phi^+$ is eaten by $W^+$, while $W^+$ decays to $\nu e^+$ or $u d^c$, a potential decay of dark matter need not consider these bosons in the product, similar to the previous model. So if dark matter decays, the final state only includes $(e,u,d)$ plus others that are trivial under $\mathcal{Z}_4$ and neutral under the electric and color charges. On the other hand, dark matter that transforms as $\underline{1}^{(2)}$ or $\underline{1}^{(4)}$ cannot decay to usual particles, since it is odd under $t^2_S=g_S$, thus suppressed by the $t^2_S$ conservation, similar to the previous model. Dark matter that transforms as $\underline{1}^{(3)}$ cannot decay to a state that contains a single field of $e,u,d$, because of the electric and color charge conservation. Also, it cannot decay to a state that contains any two of $e,u,d$, since the charge conservation restricts such state to $e^+ e^-$, $u^c u$, and $d^c d$, but this is not allowed by $\mathcal{Z}_4$. Moreover, a final state that includes three of them is only $e u d^c$ or $udd$, but not allowed by $\mathcal{Z}_4$. And, a final state that has four of them is only $euud$, $e^+ ddd$, or others as combined of two states of two $(e,u,d)$ fields above, but discarded by $\mathcal{Z}_4$. Finally, a final state that contains more than four of them must be $eeuuu$ or combine the above states due to the charge conservation, which are all suppressed by $\mathcal{Z}_4$. Hence, $SU(3)_C$, $U(1)_Q$, and $\mathcal{Z}_4$ do not permit any dark matter decaying to ordinary particles. Notice that the dark fields may have self-interactions; for example, the case of the simplest dark fields above includes a Lagrangian term, $\Phi_3^\dag\Phi_2^2$. The two dark fields are simultaneously stabilized responsible for two-component dark matter, if the mass of $\Phi_3$ is smaller than twice that of $\Phi_2$.

\subsection{The model with $\delta=1$}

From the last column of Table \ref{5RD}, we derive that $R_D = 1$ for all fields with the minimal value of $|k| = 6$, except for $k = 0$. Hence, the residual symmetry is automorphic to
\be Z_6=\lbrace 1,p,p^2,p^3,p^4,p^5\rbrace,\ee 
where $p\equiv e^{i\pi D} = (-1)^D$ and $p^6=1$. Following \cite{VanDong:2020cjf,VanDong:2020bkg}, 
we factorize $Z_6$ into
\be Z_6\cong Z_2 \otimes Z_3, \ee
where $Z_2$ is the invariant subgroup of $Z_6$, while $Z_3$ is the quotient group of $Z_6$ by $Z_2$, given by 
\bea Z_2 &=&\lbrace 1, p^3\rbrace, \\
Z_3 &=&\lbrace [1],[p^2],[p^4]\rbrace.\eea Here each (coset) element of $Z_3$, denoted by $[y]$, contains two elements of $Z_6$, the characteristic $y$ and the other $p^3 y$ as multiplied by $p^3$, thus $[1]=[p^3]=\lbrace 1,p^3\rbrace$, $[p^2]=[p^5]=\lbrace p^2,p^5\rbrace$, and $[p^4]=[p]=\lbrace p,p^4\rbrace$ are well understood. 

Since $[p^4]=[p^2]^2=[p^2]^*$ and $[p^2]^3=[1]$, $Z_3$ is completely generated by the element, 
\be
[p^2]=[e^{i2\pi D}]=[\omega^{3D}],\label{dttn1}
\ee where $\omega\equiv e^{i2\pi/3}$ defines the cube root of unity. Whereas, $Z_2$ is generated by the element,
\be p^3=e^{i3\pi D}=(-1)^{3D}.\label{dttn2}\ee It is noted that $3D$ is always integer, since $p^6=1$.

The theory conserves both the residual symmetries, $Z_2$ and $Z_3$, after symmetry breaking. Notice that $Z_2$ has two one-dimensional irreducible representations, $\underline{1}^{(1)}$ according to $p^3=1$ and $\underline{1}^{(2)}$ according to $p^3=-1$, whereas $Z_3$ has three one-dimensional irreducible representations, $\underline{1}^{(1)}$ according to $[p^2]=[ 1 ]= \{1, 1\}$ or $\{1, -1\}\rightarrow 1$, $\underline{1}^{(2)}$ according to  $[p^2]=[\om ]= \{\omega, \omega\}$ or $\{\omega, -\omega\}\rightarrow \om $, and $\underline{1}^{(3)}$ according to $[p^2]=[\om^2]=\{\omega^2, \omega^2\}$ or $\{\omega^2, -\omega^2\}\rightarrow \om^2$. Here, the representations of $Z_3$ are independent of the values $p^3=\pm 1$ that identify those of $Z_6$ in a coset; hence, the representations of $Z_3$ are determined by a homomorphism of $Z_6$ representations to those of $Z_3$, such as $[p^2]=[r]=\{r,\pm r\}\rightarrow r$, corresponding to $\underline{1}^{(1)}$, $\underline{1}^{(2)}$, and $\underline{1}^{(3)}$ for $r=1$, $\om$, and $\om^2$, respectively.

The representations of all fields under the $Z_{2,3}$ groups are computed, given in Table \ref{Z2Z3}. It is clear that the $\underline{1}^{(3)}$ representation of $Z_3$ is not presented for the existing fields. But, let us note that the antiquarks transform as $\underline{1}^{(3)}$, since $\underline{1}^{(2)*}=\underline{1}^{(3)}$, and vice versa. 
\begin{table}[h]
\bc
\caption{Field representations under $Z_6\cong Z_2\otimes Z_3$.}
\label{Z2Z3}  
\begin{tabular}{lrlc}
\hline\noalign{\smallskip}
Field & $p^3$ & $[p^2]=\{p^2,p^5\}$ & $Z_2\otimes Z_3$ \\
\noalign{\smallskip}\hline\noalign{\smallskip}
$\nu$ & $-1$ & $\{1,-1\}\rightarrow 1$ & $\underline{1}^{(2)}\otimes\underline{1}^{(1)}$ \\
$e$ & $1$ & $\{1,1\}\rightarrow 1$ & $\underline{1}^{(1)}\otimes\underline{1}^{(1)}$ \\
$u$ & $-1$ & $\{\omega,-\omega\}\rightarrow \om$ & $\underline{1}^{(2)}\otimes\underline{1}^{(2)}$ \\
$d$ & $1$ & $\{\omega,\omega\}\rightarrow \om $ & $\underline{1}^{(1)}\otimes\underline{1}^{(2)}$ \\
$\chi$ & $1$ & $\{1,1\}\rightarrow 1$ & $\underline{1}^{(1)}\otimes\underline{1}^{(1)}$ \\
$\phi^+,W^+$ & $-1$ & $\{1,-1\}\rightarrow 1$ & $\underline{1}^{(2)}\otimes\underline{1}^{(1)}$ \\
$\phi^0,A$ & $1$ & $\{1,1\}\rightarrow 1$ & $\underline{1}^{(1)}\otimes\underline{1}^{(1)}$ \\ 
\noalign{\smallskip}\hline
\end{tabular}
\ec
\end{table}

Furthermore, because the spin parity is always conserved, we conveniently multiply the residual symmetry with the spin parity group, to form \be Z_6\otimes S\cong (Z_2\otimes S)\otimes Z_3.\ee Here $Z_2\otimes S$ contains an invariant subgroup, 
\be P=\lbrace 1, P_M\rbrace,\ee  
with 
\be P_M=p^3\times P_S=(-1)^{3D+2s}.\ee Hence, we have 
\be Z_6\otimes S\cong [(Z_2\otimes S)/P]\otimes P \otimes Z_3.\ee Since $(Z_2\otimes S)/P=\{P, \{p^3,P_S\}\}$ is conserved if $P_M$ (or $P$) is conserved, we consider the remaining factor,   
\be Z_6\otimes S\supset P\otimes Z_3 \ee
to be the relevant residual symmetry instead of $Z_6$. The quotient group, $Z_3$, and its representations are remained, while $P$ has two one-dimensional irreducible representations, $\underline{1}^{(1)}$ according to $P_M=1$ and $\underline{1}^{(2)}$ according to $P_M=-1$. Thus, representations of all fields under $P$ and $Z_3$ are (re)supplied in Table \ref{PZ3}.
\begin{table}[h]
\bc
\caption{Field representations under $P\otimes Z_3$.}
\label{PZ3}  
\begin{tabular}{lrlc}
\hline\noalign{\smallskip}
Field & $P_M$ & $[p^2]$ & $P\otimes Z_3$\\
\noalign{\smallskip}\hline\noalign{\smallskip}
$\nu$ & $1$ &  $1$ & $\underline{1}^{(1)}\otimes\underline{1}^{(1)}$\\
$e$ & $-1$ & $1$ & $\underline{1}^{(2)}\otimes\underline{1}^{(1)}$\\
$u$ & $1$ & $\omega$ & $\underline{1}^{(1)}\otimes\underline{1}^{(2)}$\\
$d$ & $-1$ & $\omega$ & $\underline{1}^{(2)}\otimes\underline{1}^{(2)}$\\
$\chi$ & $1$ & $1$ & $\underline{1}^{(1)}\otimes\underline{1}^{(1)}$\\
$\phi^+,W^+$ & $-1$ & $1$ & $\underline{1}^{(2)}\otimes\underline{1}^{(1)}$\\
$\phi^0,A$ & $1$ & $1$ & $\underline{1}^{(1)}\otimes\underline{1}^{(1)}$\\ 
\noalign{\smallskip}\hline
\end{tabular}
\ec
\end{table}

From Table \ref{PZ3}, we see that the model with $\delta=1$ provides three kinds of dark field, in which the first kind transforms nontrivially under $P$ and trivially under $Z_3$, the second kind transforms trivially under $P$ and nontrivially under $Z_3$, and the last kind transforms nontrivially under both $P$ and $Z_3$. Therefore, these kinds of dark field are determined in Table \ref{darkfieldZ6} as $\Psi_{5,6,7}$ fields, respectively, where $d_{5,6,7}$ are integer. 
\begin{table}[h]
\bc
\caption{Distinct kinds of dark field implied by $P\otimes Z_3$.}
\label{darkfieldZ6}  
\begin{tabular}{lcccc}
\hline\noalign{\smallskip}
Field & $P\otimes Z_3$ & Fermion & Scalar\\
\noalign{\smallskip}\hline\noalign{\smallskip}
$\Psi_5$ & $\underline{1}^{(2)}\otimes\underline{1}^{(1)}$ & $(1,1,0,2d_5)$ & $(1,1,0,2d_5+1)$\\
$\Psi_6$ & $\underline{1}^{(1)}\otimes\underline{1}^{(2)}$ & $(1,1,0,2d_6\pm 1/3)$ & $(1,1,0,2d_6\pm 2/3)$\\
$\Psi_7$ & $\underline{1}^{(2)}\otimes\underline{1}^{(3)}$ & $(1,1,0,2d_7\pm 2/3)$ & $(1,1,0,2d_7\pm 1/3)$\\
\noalign{\smallskip}\hline
\end{tabular}
\ec
\end{table}

Among these solutions, we assume the simplest dark matter candidates corresponding to the most minimal $N$ charges, summarized in Table \ref{canZ6}. The model responsible for two-component dark matter will be based upon $F_5$ and $F_6$, which are self-interacted through a heavier dark field, $\Phi_7$. The necessary condition for $F_{5}$ and $F_{6}$ to be co-stabilized is that the net mass of $F_5$ and $F_6$ must be smaller than that of $\Phi_7$, i.e. $m_5+m_6<m_7$, where $m_{5,6,7}$ are the mass of $F_{5,6}$ and $\Phi_7$, respectively. 
\begin{table}[h]
\bc
\caption{Simplest dark matter candidates implied by $P\otimes Z_3$.}
\label{canZ6}  
\begin{tabular}{lcccc}
\hline\noalign{\smallskip}
Field & Spin & $P_M$ & $[p^2]$ & $P\otimes Z_3$ \\
\noalign{\smallskip}\hline\noalign{\smallskip}
$F_5\sim (1,1,0,0)$ & $1/2$ & $-1$ & 1 & $\underline{1}^{(2)}\otimes\underline{1}^{(1)}$ \\
$F_6\sim (1,1,0,1/3)$ & $1/2$ & 1 & $\om$ & $\underline{1}^{(1)}\otimes\underline{1}^{(2)}$ \\
$\Phi_7\sim (1,1,0,-1/3)$ & $0$ & $-1$ & $\om^2$ & $\underline{1}^{(2)}\otimes\underline{1}^{(3)}$ \\
\noalign{\smallskip}\hline
\end{tabular}
\ec
\end{table}

The phenomenology of this model is completely different from the $U(1)_{B-L}$ extension \cite{VanDong:2020bkg}, since the ordinary particles transform nontrivially under $P_M$ in the same with the dark fields (cf. Table \ref{PZ3}). Additionally, it happens similarly for $Z_3$, where the quarks are nontrivial as the dark fields. The electric and color charge conservation is  necessarily to ensure dark matter stability. And, the dark matter components can have arbitrary masses, which need not be smaller than the $e,u,d$ masses. Let us see.  

Since $P_M$ is identical to $g_S$, the stability of the dark matter component $F_5$ is always ensured by the $SU(3)_C$, $U(1)_Q$, and $P$ symmetries, independent of its mass, analogous to the case of the dark matter $\Psi_1$ in the model with $\delta=-1$. Whereas, the $SU(3)_C$ and $Z_3$ symmetries are responsible for the stability of the dark matter component $F_6$, where only quarks $u,d$ transform nontrivially under $Z_3$ as $F_6$. Prove: Since $F_6$ as a dark matter is color neutral, it cannot decay to any colored state, such as those that include a single quark. The $SU(3)_C$ conservation requires a color-neutral final state if resulting from a decay of $F_6$, by assumption. Hence, this final state containing quarks (otherwise, suppressed by $Z_3$) must be composed of $q^c q$ and/or $qqq$. However, these combinations are trivial under $Z_3$, implying that the final state is invariant under $Z_3$ which cannot be a product of $F_6$ decay to be a contradiction. In other words, the $SU(3)_C$ and $Z_3$ symmetries suppress the decay of $F_6$, even if $F_6$ has a mass larger than that of quarks.

\section{\label{pheno2c} Phenomenology}

\subsection{Neutrino mass generation}

The charged leptons and quarks gain appropriate masses through the Yukawa couplings with the usual Higgs field $\phi$, similar to the standard model. 

However, the advantage is relevant to the generation of neutrino masses. Indeed, the neutrinos $\nu_{L,R}$ possess the Yukawa terms,
\be \mathcal{L} \supset h^\nu_{ab}\bar{l}_{aL}\tilde{\phi} \nu_{bR}+\fr 1 2 f^\nu_{ab}\bar{\nu}^c_{aR} \nu_{bR}\chi+H.c.\ee 
Substituting the VEVs, $\nu_{R}$ obtain a Majorana mass matrix, $m_R=-f^\nu\La/\sqrt{2}$, while $\nu_{L,R}$ receive a Dirac mass matrix, $m_D=-h^\nu v/\sqrt{2}$. They are given through the mass Lagrangian,
\bea \mathcal{L}\supset -\fr 1 2 \left(\bar{\nu}_{aL}\ \bar{\nu}^c_{aR}\right)
\left(\begin{array}{cc}
0 & (m_D)_{ab}\\
(m_D)_{ba} & (m_R)_{ab}\end{array}\right)
\left(\begin{array}{c}
\nu^c_{bL}\\
\nu_{bR}\end{array}\right)+H.c.\eea 

Since $v\ll \La$, we have $m_D\ll m_R$. Hence, the observed neutrino ($\sim \nu_L$) masses are given by the seesaw mechanism, 
\be m_\nu\simeq -m_D m^{-1}_R m^T_D = h^\nu (f^\nu)^{-1} (h^\nu)^T \fr{v^2}{\sqrt{2}\La},\ee while the heavy neutrinos ($\sim \nu_R$) have a mass proportional to $m_R$ at $\La$ scale. 

The breaking of the hyperdark charge $N$ is necessarily to generate small neutrino masses, proportional to $m_\nu\sim v^2/\La$, where $\La$ is large, compared with the weak scale. This breaking (or seesaw) scale, $\La$, is constrained by the particle colliders below.  

\subsection{Collider searches for $Z'$}

The $U(1)_N$ gauge boson, $Z'$, obtains a mass, \be m_{Z'}\simeq 2 |\delta|g_N\La,\ee
where $g_N$ is the $U(1)_N$ gauge coupling, and that small contributions due to mixing effects from the kinetic mixing term and the weak breaking are neglected. 

The field $Z'$ couples to the usual fermions and $\nu_R$ via the Lagrangian, \be \mathcal{L} \supset -g_N [N({f_L}) \bar{f}_L \ga^\mu f_L+N({f_R}) \bar{f}_R \ga^\mu f_R] Z'_\mu \equiv \bar{f}\ga^\mu[g^{Z'}_V(f)-g^{Z'}_A(f)\ga_5]f Z'_\mu,\ee where we define
\be g^{Z'}_V(f)=-\fr 1 2 g_N [N(f_L)+N(f_R)],\hs g^{Z'}_A(f)=-\fr 1 2 g_N [N(f_L)-N(f_R)].\label{dtsn111}\ee This applies for usual fermions, except for neutrinos. By the seesaw mechanism, the usual neutrinos $\nu\simeq \nu_L$ and the sterile neutrinos $\nu'\simeq \nu_R$ are distinct (Majorana) particles. Hence, $g^{Z'}_{V,A}(\nu)$ are computed by suppressing $N(\nu_R)$, whereas $g^{Z'}_{V,A}(\nu')$ are achieved by omitting $N(\nu_L)$.   

If $m_{Z'}$ is smaller than the highest collision energy of the LEPII experiment, $\sqrt{s}=209$ GeV, the particle $Z'$ may be resonantly produced at the LEPII. The agreement between the experimental measurement and the standard model indicates that $g_N\lesssim 10^{-2}$, up to a factor of $\delta$ \cite{Appelquist:2002mw}. 

If $m_{Z'}>209$ GeV, the process $e^+e^-\to f\bar{f}$ gets an off-shell contribution of $Z'$, described by the effective Lagrangian, \bea \mathcal{L}_{\mathrm{eff}} &\supset& \fr{g^2_N N({e_L})N({f_L})}{m^2_{Z'}}(\bar{e}_L \ga^\mu e_L)(\bar{f}_L\ga_\mu f_L)\crn
&&+(LR)+(RL)+(RR).\eea Taking $f=\mu,\tau$, the LEPII searched for such chiral structures, yielding a strong limit for $Z'$ mass per coupling to be \be \fr{m_{Z'}}{|\delta-1/2|g_{N}}\geq 6\ \mathrm{TeV}, \ee which translates to $\La\geq 1.5|(2\delta-1)/\delta|$ TeV \cite{Carena:2004xs}. 

The LHC searched for dilepton signals through $pp\to f\bar{f}$ for $f=e,\mu,\tau$, mediated by $Z'$, which provide a mass bound about $m_{Z'}=4$ TeV for $Z'$ coupling similar to $Z$. This indicates to a bound for $\La\sim m_{Z'}/2g\sim 3$ TeV, where $g$ is the $SU(2)_L$ coupling \cite{Aaboud:2017buh}. 

All the above searches imply a bound on the new physics scale to be $\La\sim m_{Z'}/g_N>\mathcal{O}(1)$ TeV. Note that $Z'$ has not been considered, decaying to the dark fields. By contrast, if $Z'$ decays to some dark fields, such bound is more relaxed. 

Let us remind the reader that although $\La$ is constrained at TeV, the $Z'$ mass, $m_{Z'}\sim g_N \La$, may be small, as $g_N$ is small, and vice versa. 

\subsection{$U(1)_N$ running coupling}

Notice that the interactions of dark fields with $U(1)_N$ gauge portal, as well as dark field self-couplings, govern the dark matter observables. Here, such interaction of a dark field, called $X$, with $Z'$ is given by the Lagrangian, either $\mathcal{L}\supset \bar{X} (i\ga^\mu D_\mu - m_X)X$ or $\mathcal{L}\supset (D^\mu X)^\dagger (D_\mu X)-m^2_X X^\dagger X$, for $X$ to be a dark fermion or scalar, respectively, where $D_\mu=\pa_\mu + i g_N N_X Z'_\mu$ is the covariant derivative, and $N_X$ is the hyperdark charge of the dark field. Due to the cyclic property of discrete symmetry, $N_X$ may be arbitrary, similar to the ones given via $d_{1,2,\cdots,7}$ above. One may wonder that the theory is not well defined for large $N_X$.

First of all, opposite to non-Abelian charges, Abelian charges are completely arbitrary, as also mentioned from outset. Additionally, the Lagrangian of a $U(1)_N$ gauge theory always conserves a scaling symmetry, \be g_N\rightarrow g'_N=g_N/c,\hs N\rightarrow N'=c N,\ee where $g'_N$ and $N'$ are the coupling and charge after transformation by any value of $c$. Thus, one may work in a basis, such that $g_N$ is infinitesimal, while $N$ is rarely big, leaving the physics unchanged. Further, the running of $g_N$ along with energy scale $\mu$ satisfies the RG equation, \be \mu \pa g_N/\pa \mu = \beta(g_N) = -(g^3_N/16\pi^2) b_N,\ee where the beta function, \be b_N=-\fr 2 3 \sum_L N^2_L-\fr 2 3 \sum_R N^2_R-\fr 1 3 \sum_S N^2_S,\ee is summed over the left and right chiral fermions and scalars, respectively. The RG equation preserves the above scaling symmetry as a result. This implies that a theory with large $N$ behaves similarly to the normal ones, when the energy scale slides. A special feature of the Abelian theory is $b_N<0$ for every $N$. Hence, the coupling $g_N$ decreases when $\mu$ decreases. Given that the theory is well defined at the Planck or grand unification scale, i.e. $g_N N\sim 1$, it works perturbatively and is predictive below such a large scale.\footnote{An exception is that the hyperdark charge may not be unified with the normal charges in a GUT framework, since it necessarily results from a dequantization of electric charge. However, such charge possibly originates from a string \cite{Faraggi:2011xu} or a flipped GUT~\cite{Huong:2016kpa}.}  
 
As a consequence, we divide the dark matter phenomenology into two cases. If $g_N$ is small, the relevant dark matter may have a large hyperdark charge. This would imply a large signal strength for dark matter when scattering with electrons, as shown  below (see also \cite{VanDong:2020bkg} for $B-L$ model). By contrast, if $g_N$ is large, the charge $N$ should be small as of usual particles. In this case, the $Z'$ portal effectively governs the WIMP components, co-existed beyond the weak scale, also shown below (see also \cite{Nam:2020twn} for noncommutative $B-L$ model).    

\subsection{\label{XENON1T} Small $U(1)_N$ coupling: Probing the XENON1T excess}

Recently the XENON1T collaboration announced an excess in electronic recoil energy ranging from 1 to 7 keV, peaked about 2.3 keV, with a statistical signification above 3$\sigma$, with 285 events produced above the expected background of $232 \pm 15$ events \cite{Aprile:2020tmw}. 

This excess may be manifestly explained within the scenarios of two-component dark matter obtained in this work, given that one dark matter component is boosted with a velocity of $0.1c$ order~\cite{Kannike:2020agf,Fornal:2020npv,Primulando:2020rdk,Su:2020zny,Cao:2020bwd,Alhazmi:2020fju,DelleRose:2020pbh,Ko:2020gdg,Dey:2020sai}. Additionally, the $U(1)_N$ coupling constant, $g_N$, is small enough to evade the low energy experimental bounds, simultaneously this enhances the charge of boosted dark matter, thus its signal strength as measured by the XENON1T \cite{VanDong:2020bkg}. 

In the model with $\delta=2$, both the dark matter components $\Phi_{2,3}$ populate our galaxy, possessing densities set by their annihilation to the usual leptons and quarks through the $U(1)_N$ gauge boson portal ($Z'$), as well as the self-annihilation between them by the self-couplings, $\Phi^\dagger_3 \Phi^2_2$ and $(\Phi^\dagger_2 \Phi_2)(\Phi^\dagger_3 \Phi_3)$. If $\Phi_{2,3}$ have masses, called $m_{2,3}$, respectively, such that $m_2 < m_3 < 2m_2$, $\Phi_3$ is boosted in annihilation $\Phi_2 \Phi_2 \rightarrow \Phi_3 Z'$, while $\Phi_2$ is boosted in annihilation $\Phi_3\Phi_3\rightarrow \Phi_2\Phi_2$. Whereas, if $m_3<m_2$, only the component $\Phi_3$ is boosted through the annihilations, $\Phi_2\Phi_2\rightarrow \Phi_3\Phi_3, \Phi_3 Z'$. Above, $Z'$ has a mass to be smaller than those of $\Phi_{2,3}$ due to the small $g_N$. The boosted dark matter is either both $\Phi_{2,3}$ or only $\Phi_3$, depending on their relative masses. 

In the model with $\delta=1$, the dark matter component $F_5$ contributes to the whole cold dark matter, set by its annihilation to $F_6$ via the self-interaction $F_5 F_6\Phi_7$, where $m_6<m_5$. The component $F_6$ has vanished relic density due to its annihilations to $Z'$ as well as the usual particles. However, $F_6$ is presently produced, boosted in the annihilation of the cold dark matter $F_5$ due to the $F_5 F_6 \Phi_7$ self-interaction, similar to the thermal process. In this model, only the boosted dark matter component is $F_6$. 

In all of the mentioned models, the boosted dark matter subsequently scatters on electrons in the XENON1T experiment, causing the observed effect, as enhanced by the large charge of the boosted dark matter. The next phenomenology happens quite the same. Hence, we take only the model with $\delta=1$ into account, hereafter. 

\begin{table}[h]
\bc
\caption{Couplings of $Z'$ with usual fermions, given through Eq. (\ref{dtsn111}).}  
\label{Zp}
\begin{tabular}{lrr}
\hline\noalign{\smallskip}
$f$ & $g^{Z'}_V(f)$ & $g^{Z'}_A(f)$\\ 
\noalign{\smallskip}\hline\noalign{\smallskip}
$\nu_e,\nu_\mu,\nu_\tau$ & $-g_N/4$ & $-g_N/4$ \\
$e,\mu,\tau$ & $-g_N/4$ & $-g_N/4$ \\
$u,c,t$ & $-g_N/12$ & $g_N/4$ \\
$d,s,b$ & $5g_N/12$ & $-g_N/4$ \\
\noalign{\smallskip}\hline
\end{tabular}
\ec
\end{table}

The Lagrangian of the considering model includes \bea
\mathcal{L} &\supset& \bar{F}_5 (i\ga^\mu \pa_\mu -m_5)F_5 + \bar{F}_6(i\ga^\mu D_\mu -m_6) F_6 \crn
&&+ (D^\mu \Phi_7)^\dagger (D_\mu \Phi_7)-m^2_7 \Phi^\dagger_7\Phi_7+ \bar{e} (i \ga^\mu D_\mu - m_e)e \crn
&&+ (y\bar{F}_5F_6\Phi_7+H.c.),
\eea
where $D_\mu=\pa_\mu + i g_N N Z'_\mu$ is the covariant derivative. Note that $F_{5,6}$ are vectorlike, while the electron $e$ is chiral (cf. Table \ref{Zp}), and that the dark field masses obey $m_7>m_5+m_6$ and $m_5>m_6$, as mentioned. We have fixed $F_5\sim (1,1,0,0)$, thus it has only self-couplings to other dark fields and gravity interaction with normal matter, as desirable. Whereas, $F_6$ and $\Phi_7$ may possess generic hyperdark charges, i.e. \be F_6\sim (1,1,0,1/3+2n),\hs  \Phi_7\sim (1,1,0,-1/3-2n),\ee for $n$ integer, which differ from the minimal basic charges in Table~\ref{canZ6} by even numbers, similar to the ones in Table~\ref{darkfieldZ6}. This is always allowed by the cyclic property of residual $Z_3$ and $P$ groups and obviously that they do not change the representations of the fields. 

\begin{figure}[h]
\bc
\resizebox{0.6\textwidth}{!}{%
  \includegraphics{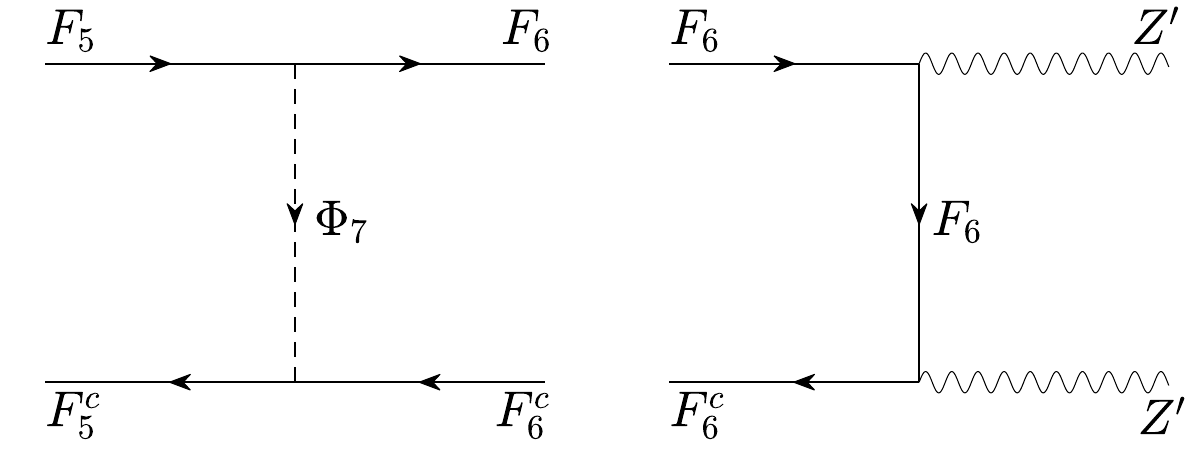}}
\caption{Dark matter conversion and annihilation of the second dark matter. The second process has an extra $u$-channel, but undepicted, for simplicity.}
\label{darkcharge1}      
\ec
\end{figure}

\begin{figure}[h]
\bc
\resizebox{0.35\textwidth}{!}{%
  \includegraphics{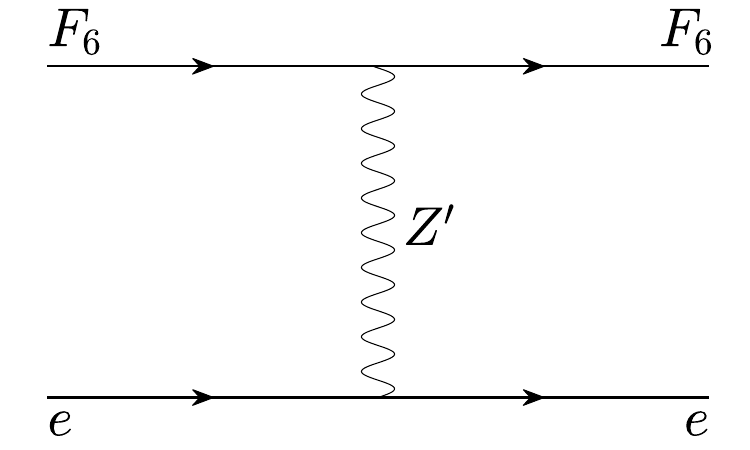}}
\caption{Scatter of the boosted dark matter with electron.} 
\label{xxenone}
\ec     
\end{figure}

The cold dark matter $F_5$ obtains a correct density by its $t$-channel annihilation to the second dark matter, i.e. $F_5 F_5\to F_6 F_6$, exchanged by the $\Phi_7$ dark field, as depicted by the left diagram in Fig. \ref{darkcharge1}. $F_6$ strongly couples to the $U(1)_N$ gauge boson $Z^\prime$, by which it has a vanished density given by the right diagram in Fig. \ref{darkcharge1}. Note that $F_6$ also annihilates to the usual particles via $Z'$ portal, but it is small, compared to the given channel to $Z'$, since its hyperdark charge is large. The densities of $F_{5,6}$ are explicitly computed at the end of this section. Presently, $F_6$ is boosted via a process similar to the left diagram in Fig. \ref{darkcharge1}, and subsequently this fast dark matter scatters on electrons in the XENON1T experiment via the diagram in Fig. \ref{xxenone}, by the $Z^\prime$ portal. The mildly boosted (i.e., $v\sim 0.1c$) phenomenon happens when $m_5\approx m_6$, and we also assume $m_{5,6}> m_e$ to satisfy the BBN and CMB bounds \cite{Sabti:2019mhn}. 

By the process in Fig. \ref{xxenone}, the electronic recoil energy approaches $E_R < 2 m_e v^2$ for $m_{6}>m_e$ and agrees with the XENON1T if $F_6$ is boosted with $v^2=(m_5^2-m^2_6)/m^2_5\sim1\%$ \cite{Kannike:2020agf}. The transferred momentum is $|q|=E_R/v\sim 40$ keV. In the limit of the $Z'$ mass to be much larger than $|q|$, the $F_6$-$e$ scattering cross section is given by  \cite{Joglekar:2019vzy}
\be
\sigma_e=\frac{g_N^4(1/3+2n)^2 m_e^2}{8\pi m_{Z^\prime}^4}.
\ee Since $m_{Z'}\sim g_N\La$, the cross-section is enhanced for large boosted dark matter charge, as expected. 

 In the literature, two potential sources that produce the flux of boosted dark matter have been discussed, namely (i) dark matter annihilation in the current galaxy center/halo and (ii) dark matter capture and their annihilation in the sun. However, the second source does not significantly contribute to the flux in our model, because the cold dark matter $F_5\sim (1,1,0,0)$ is a gauge singlet which does not interact with the sun matter, such as nuclei, at tree level. Through 1-loop effects that include both $F_6,\Phi_7$ contributions, the field $F_5$ effectively couples to $Z'$, such as \be \mathcal{L}\supset \fr{g'_N y^2}{48\pi^2m^{2}_7}\ln\fr{m^2_6}{m^2_7}\left[\bar{F}_5 \ga^\mu \pa^\nu F_5 (\pa_\mu Z'_\nu - \pa_\nu Z'_\mu)+H.c.\right],\ee with $g'_N=|1/3+2n|g_N$, in agreement to \cite{Agashe:2014yua}. This induces a spin-independent scattering cross-section of $F_5$ on nucleons via $Z'$ exchange in the sun, \be \sigma_{p,n}=\fr{1}{\pi}\mu^2_{p,n}g^2_{p,n}\left(\fr{g'_N y^2}{48\pi^2m^{2}_7}\ln\fr{m^2_6}{m^2_7}\right)^2\fr{t^2}{(m^2_{Z'}-t)^2} \lesssim 10^{-47}\mathrm{cm}^2,\ee where we take $m_7\gtrsim 1$ GeV, the log function is of one order, and the squared momentum transfer is $t=-2m_{p,n}E_R \sim -m^2_{Z'}$, since the recoil energy typically $E_R<2m_{p,n}v^2\sim 1$ keV and $m_{Z'}\sim 1$ MeV, as shown below. Additionally, we have the reduced mass $\mu_{p,n}=m_{p,n} m_5/(m_{p,n}+m_5)\lesssim 1$ GeV, the $Z'$ vector couplings to nucleons $g_{p(n)}=1(3)g_N/4\sim 10^{-7}$ (see below for $g_N$ value and $g'_N\simeq 1$), and $y \lesssim \mathcal{O}(1)$ due to the perturbative limit. The sun will reach a capture-annihilation equilibrium, such that the scattering cross-section sets the flux of boosted dark matter to be \cite{Fornal:2020npv} \be \Phi^{\mathrm{Sun}}_{F_6}=7.2\times 10^{-9}\mathrm{cm}^{-2}\mathrm{s}^{-1}\left(\fr{\sigma_{p,n}}{10^{-47}\mathrm{cm}^2}\right)\left(\fr{1\ \mathrm{GeV}}{m_5}\right)^2,\ee which is much smaller than the following galaxy source and is neglected. 

Hence, the dark matter is not captured/accumulated in the sun core over time, i.e. the capture-annihilation phenomenon is not enhanced by the sun. The boosted dark matter flux in our model only comes from the first source (i). We assume that the dark matter obeys an NFW profile \cite{Navarro:1995iw}. The $F_6$ flux from full sky is given by \cite{Agashe:2014yua}\be \Phi_{F_6}=1.6\times 10^{-4}\mathrm{cm}^{-2}\mathrm{s}^{-1}\left(\fr{\langle \sigma v_{\mathrm{rel}} \rangle_{F_5 F_5\to F_6 F_6}}{5\times 10^{-26}\mathrm{cm}^{3}\mathrm{s}^{-1}}\right)\left(\fr{1\ \mathrm{GeV}}{m_{5}}\right)^2,\label{them11}\ee where $\langle \sigma v_{\mathrm{rel}} \rangle_{F_5 F_5\to F_6 F_6}$ is the thermally-averaged dark matter annihilation cross-section at present time, computed below in (\ref{anni56}), and 1 GeV belongs to the dark matter mass range investigated, but not necessarily indicated to a benchmark. The flux of the fast dark matter $F_6$ that comes from all the sky due to the $F_5$ cold dark matter annihilations in our galaxy would cause a signal rate at the XENON1T detector. The total number of signal events can be estimated as \be N_{\mathrm{sig}}=Z_{\mathrm{eff}}\times n_{\mathrm{Xe}}V\times T\times \sigma_e\times \Phi_{F_6},\label{them12}\ee in agreement to \cite{Fornal:2020npv}. Here, $Z_{\mathrm{eff}}\sim40$ is the effective number of recoil electrons in a Xenon atom, which is governed by electrons on $N,O$ shells; $n_{\mathrm{Xe}} V=M/m_{\mathrm{Xe}}$ is the number density of Xenons in the detector, multiplied by the fiducial volume of the detector, which equals the detector-to-Xenon mass ratio; $T$ is the total operation time; and notice that $MT=0.65$ tonne-years relevant to the current XENON1T data. 

 As shown below, the present-day dark matter is dominated by $F_5$, set by its dominant annihilation to $F_6$ in the early universe, hence constraining the cross-section $\langle \sigma v_{\mathrm{rel}} \rangle_{F_5 F_5\to F_6 F_6}\simeq 5\times 10^{-26}\mathrm{cm}^{3}\mathrm{s}^{-1}$ for the correct abundance. With this condition, from Eqs. (\ref{them11}) and (\ref{them12}), the relation between the number of signal events and the scattering cross section is derived to be 
\be
\frac{N_{\mathrm{sig}}}{100}\simeq \frac{16\sigma_e}{3\ \mathrm{pb}}\left(\frac{1\ \mathrm{MeV}}{m_5}\right)^2.
\ee The dark matter mass satisfies $m_5>m_e$ and it can pick up values up to GeV scale; above the 1 MeV conveniently put does not mean a benchmark value.  
Noting that $m_{Z^\prime}\simeq 2g_N\Lambda$ for the model with $\delta=1$, we deduce
\be
\frac{N_{\text{sig}}}{100}\simeq 6.4 \times 10^{-14} \left(\fr 1 3+2n\right)^2\left(\fr{m_e}{m_5}\right)^2\left(\frac{3\text{ TeV}}{\Lambda}\right)^4.\label{xenon}
\ee

Required to explain the number of signal events, $N_{\mathrm{sig}}\sim 100$, the hyperdark charge of dark matter is large, \be |n|\sim 1.98\times 10^6\left(\fr{m_5}{m_e}\right)\left(\fr{\La}{3\ \mathrm{TeV}}\right)^2>1.98\times 10^6,\label{boundn}\ee where both $\Lambda$ and $m_5$ have a lower bound, $\Lambda> 3$ TeV and $m_5> m_e$, as mentioned. Additionally, from Eq. (\ref{boundn}), the dark matter charge required is increased, when the dark matter mass $m_5$ and/or the new physics scale $\Lambda$ are larger than the given bounds.

In this case, we approximate the beta function through $b_N\simeq -(5/3)(1/3+2n)^2$, yielding the RG equation,
\be \mu \fr{\pa g'_N}{\pa \mu} \simeq \fr{5}{48\pi^2}g'^3_N,\ee where $g'_N=|1/3+2n|g_N$ slides as the usual couplings. Defining the hyperdark charge coupling strength, $\al'_N=g'^2_N/4\pi$, we obtain the solution, \be \fr{1}{\al'_N}=\fr{1}{\al'_{NG}}-\fr{5}{6\pi}\ln\fr{\mu}{\mu_G},\ee where $_G$ indicates to those at the large (GUT or Planck) scale. Imposing a perturbative limit for the hyperdark charge interaction at the large scale, say $\al'_{NG}=1$ at $\mu_G=10^{16}$ GeV, it leads to $\al'_N\simeq 0.08$ at scale $\mu\sim 1$ MeV of interest, or in other words, 
\be |1/3+2n|g_N\simeq 1.\label{adndt}\ee Combining this equation with the condition $N_{\mathrm{sig}}=100$ required to explain the excess, we get 
\be m_{Z'}\simeq 1.81\times 10^3 (g_N/m_5)^{1/2} \text{ MeV}^{3/2}.\label{adndt1}
\ee

From Eq. (\ref{adndt}) and Eq. (\ref{boundn}), we obtain a bound for the $U(1)_N$ gauge coupling,
\be g_N\lesssim 2.52\times 10^{-7}.\label{adndt2} \ee From Eq. (\ref{adndt1}) and Eq. (\ref{adndt2}), the mass of the $Z'$ boson is correspondingly bounded by
\be m_{Z'}\lesssim 1.28 \text{ MeV},\label{adndt23} \ee since $m_5>m_e$. This bound is very close to the sensitive limit of beam dump experiments, which detect the signal of the decay $Z'\rightarrow e^+ e^-$, that requires $m_{Z'}> 2 m_e\simeq 1$ MeV \cite{Ilten:2018crw,Bauer:2018onh}. That said, $Z'$ is always viable below such mass limit since the beam dump experiments are not able to find it, in spite of what size $g_N$ is. 

Notice that $m_{Z'}$ and $g_N$ are further constrained by the neutrino-electron scattering, horizontal branch stars, and supernova 1987A experiments, similar to the $B-L$ model \cite{VanDong:2020bkg}, which are signified as follows. The neutrino-electron scattering due to the $Z'$ exchange yielding a bound $g_N\lesssim 10^{-6}$ for $m_{Z'}\sim $ MeV is obviously suppressed by the condition (\ref{adndt2}). The experiment with horizontal branch stars studies energy loss, carried by $Z'$, implying a bound $m_{Z'}>0.34$~MeV for $g_N$ satisfying (\ref{adndt2}), but being above the excluded regime $g_N\nsim 10^{-8}$--$10^{-11}$, as constrained by the supernova 1987A. The BBN supplies a lower limit on $g_N$, which would exclude the viable lower region with $g_N\lesssim 10^{-11}$, according to $m_{Z'}\sim$ MeV. In summary, in addition to the XENON1T constraints in (\ref{adndt23}) and (\ref{adndt2}), the astrophysical and cosmological experiments restrict 
\be m_{Z'} > 0.34\ \mathrm{MeV}\ \mathrm{and}\ g_N\gtrsim 10^{-8}.\label{anhtvao1}\ee Because $Z'$ is light, potentially affecting the BBN measurement, in what follows we will discuss the BBN bound, for clarity. 

Since the dark matter components $F_{5,6}$ are above the BBN bound given at MeV, we need only examine two cases of $Z'$ mass. If the $Z'$ mass is also above this BBN bound, no sub-MeV hidden states exist. Hence, no new relativistic degrees of freedom are presented during the BBN. Only constraint from the BBN is that $Z'$ decays into the standard model fast enough, larger than the Hubble rate, in order to avoid late dissociation process. Thus, the decay $Z'\to e^+e^-, \nu \nu^c$ before $T\sim$ MeV, say $\Ga(Z'\to e^+e^-,\nu\nu^c)\geq H(T\sim\mathrm{MeV})$, requires \be g_N\gtrsim 0.46\times 10^{-9}(\mathrm{MeV}/m_{Z'})^{1/2}\sim 0.46\times 10^{-9}.\label{anhtvao2}\ee If the $Z'$ mass is below the BBN bound, the field $Z'$ does not significantly contribute to the allowed number of relativistic degrees of freedom during the BBN at the standard model temperature. Indeed, after the neutrino decoupling, a significant population of $Z'$ may be generated through inverse decays, such as $\nu\nu^c\to Z'$, known as a freeze-in mechanism. Note that in this regime, a population of $Z'$ derived through the reaction $\ga e^\pm \leftrightarrow Z' e^\pm$ or $e^+e^-\leftrightarrow Z'\ga$ is ineffective, because such process is suppressed by a factor $\al\simeq 1/137$ relative to the present process $\nu \nu^c\leftrightarrow Z'$ and therefore is safely neglected.\footnote{The process $e^+ e^-\leftrightarrow Z' $ does not occur for cosmic temperatures and $m_{Z'}$ below 1 MeV, so it would not be included. Also, similar processes that thermalize $Z'$ with the light quarks $(u,d)$ are ineffective in comparison to $\nu \nu^c\leftrightarrow Z'$.} Once the $Z'$ population is in equilibrium with the neutrino population, it will maintain the distribution through the active decay $Z'\to \nu\nu^c$ and the inverse decay $\nu\nu^c\to Z'$. Until the temperature drops to $T_\nu \sim m_{Z'}/3$, at which time $Z'$ decays out of equilibrium, as the inverse decay is kinetically suppressed, the active decay $Z'\to \nu\nu^c$ potentially produces a net contribution to the effective number of neutrinos, $N_{\mathrm{eff}}$. This contribution is viable, since the decay products of $Z'$ have an energy $E_\nu\sim m_{Z'}/2$, generally larger than $3T_\nu$, implying a final neutrino distribution to be more energetic than that found in thermal equilibrium.   

The momentum distribution functions of $Z'$ and $\nu$, labeled $f_{Z'}$ and $f_\nu$ respectively, obey the Boltzmann equations, such as \cite{Kawasaki:1992kg} \bea &&\fr{\pa f_{Z'}}{\pa t}-H p_{Z'}\fr{\pa f_{Z'}}{\pa p_{Z'}}=-\fr{m_{Z'}\Ga_{Z'}}{E_{Z'}p_{Z'}}\int^{(E_{Z'}+p_{Z'})/2}_{(E_{Z'}-p_{Z'})/2}dE_\nu F(E_{Z'},E_\nu,E_{Z'}-E_\nu),\label{at123}\\
&& \fr{\pa f_{\nu}}{\pa t}-H p_{\nu}\fr{\pa f_{\nu}}{\pa p_{\nu}}=\fr{m_{Z'}\Ga_{Z'}}{E_{\nu}p_{\nu}}\int^{\infty}_{|(m^2_{Z'}/4p_\nu)-p_{\nu}|}dp_{Z'}\fr{p_Z'}{E_{Z'}} F(E_{Z'},E_\nu,E_{Z'}-E_\nu),\label{at124}\eea where $F(x,y,z)=f_{Z'}(x)[1-f_\nu (y)][1-f_\nu (z)]-f_\nu (y) f_\nu (z) [1+f_{Z'}(x)]$, $\Ga_{Z'}=g^2_N m_{Z'}/64\pi$ is the rest frame width, and $H=\dot{a}/a$ is the cosmic expansion rate with $a$ the scale factor. The energy densities of the decay particle and the products are obtained by \be \rho_{Z'}=\fr{g_{Z'}}{2\pi^2}\int_0^\infty p^2_{Z'}d p_{Z'} E_{Z'} f_{Z'},\hs \rho_\nu=\fr{g_\nu}{2\pi^2}\int_0^\infty p^2_\nu d p_\nu E_\nu f_\nu,\ee where $g_{Z'}=3$ and $g_\nu=2$ are the number of spin states of $Z'$ and $\nu$, respectively. When the temperature drops below $m_{Z'}/3$, the inverse decays are forbidden and the entropy of $Z'$ population is transferred to other species. If only neutrinos and photons exist, the effective number of neutrinos is \be N_{\mathrm{eff}}=\fr 8 7 \left(\fr{11}{4}\right)^{4/3}\fr{\rho_\nu}{\rho_\ga},\ee where $\rho_\nu$ differs from the SM value due to the entropy transferred from $Z'$ decays. The deviation of the effective number is \be \Delta N_{\mathrm{eff}}=\fr{360}{7\pi^4}\int_0^\infty dy_\nu y^3_\nu (f_\nu-f^\mathrm{FD}_\nu),\ee which is integrated over comoving momentum of $\nu$, $y_\nu =a p_\nu /\mathrm{MeV}$, where $f_\nu^\mathrm{FD}=(1+e^{y_\nu})^{-1}$ is the distribution function of a free-streaming decoupled neutrino. The numerical investigation of the Boltzmann equations (\ref{at123}) and (\ref{at124}) in the fast decay regime $\Ga_{Z'}\geq H(m_{Z'})$ was already done in \cite{Escudero:2019gzq}, with noticing that at very high temperatures $f_{Z'}=0$, and its result would apply to our model without change. That said, the deviation of the effective number of neutrinos is $\Delta N_{\mathrm{eff}}\simeq 0.21$. This value is suitable to the Hubble tension between local measurements and temperature anisotropies of CMB, reported in \cite{Riess:2016jrr}. Notice that the fast decay regime $\Ga(Z'\to \nu \nu^c)\geq H(m_{Z'})$ under consideration requires \be g_N\gtrsim 0.53\times 10^{-9}(m_{Z'}/\mathrm{MeV})^{1/2}\sim 0.53\times 10^{-9},\label{daccdtt}\ee different from the previous case of the $Z'$ mass. Hence, the new light state $Z'$ may be presented, which does not alter the BBN measurement. 

It is noted that $F_{5,6}$ are populated and follow a thermal distribution before the freeze-out of the dark matter. Hence, $Z'$ must also be populated, since it is annihilated in by $F_6$, and should have enough interactions with the standard model sector to keep the dark matter in thermal equilibrium until freeze-out. However, the freeze-in mechanism discussed above is valid only if $g_N$ is sufficiently small such that the $Z'$ population will not reach equilibrium with the standard model thermal bath in the very early universe, by contrast.\footnote{In this case, the dark sector temperature may be very different from that of the standard model sector, depending on their couplings to the inflaton and entropy releases, and thus the WIMP paradigm breaks down.} The condition for $Z'$ not to reach equilibrium with the heavy leptons ($\mu,\tau$) and the heavy quarks ($c,b,t$) at early times can be considered as $n_\mu \langle \sigma v\rangle_{\mu^+\mu^-\to \ga Z'}<H(m_\mu)$, because the annihilations of $\tau,c,b,t$ to $\ga Z'$ are quite smaller than that of $\mu$ due to the suppressions of larger fermion masses and/or smaller $\ga,Z'$ couplings. Correspondingly, the condition requires \be g_N\lesssim 4\left(\fr{1.66\sqrt{g_*}}{\al}\fr{m_\mu}{m_{\mathrm{Pl}}}\right)^{1/2}\sim 10^{-8},\label{daccdtt1}\ee where $\langle \sigma v\rangle_{\mu^+\mu^-\to \ga Z'}\sim \al g^2_N/16m^2_\mu$ is approximately given in non-relativistic limit near $T\sim m_\mu$ and is independent of $m_{Z'}$ due to $m_{Z'}\ll m_\mu$. From Eqs. (\ref{daccdtt}) and (\ref{daccdtt1}), there is a sizable range of $g_N$, i.e. $0.53\times 10^{-9}\lesssim g_N\lesssim 10^{-8}$, under which the $Z'$ population does not reach equilibrium with the heavy fermions at early times, but $Z'$ will equilibrate with neutrinos, leading to a $N_{\mathrm{eff}}$ deviation through its decay $Z'\to \nu\nu^c$ appropriately. Unfortunately, this range of $g_N$ that ensures the freeze-in mechanism is obviously excluded by the condition in Eq. (\ref{anhtvao1}) which comes from the supernova 1987A and is not favored by the WIMP criteria, as mentioned. Nonetheless, this study of the BBN bounds given in Eqs. (\ref{anhtvao2}) and (\ref{daccdtt}) is aimed at ruling out the bound $g_N\lesssim 10^{-11}$ as hinted from the supernova 1987A too. It is noteworthy that $g_N$ is finally bounded by Eq. (\ref{anhtvao1}) and in this case the condition in Eq. (\ref{daccdtt1}) is not satisfied. Hence, $g_N\gtrsim 10^{-8}$ ensures $Z'$ to be in kinetic equilibrium with the thermal plasma at early times, protecting the freeze-out mechanism of the dark matter from outset, and this dark matter setup also independently excludes $g_N\lesssim 10^{-11}$.        

Last, but not least, let us evaluate the dark matter relic densities as stated. It is suitably to assume $m_{6}>m_{Z'}$, as above mentioned. Hence, $F_6$ is obviously annihilated to $Z'$ via the right diagram in Fig. \ref{darkcharge1}. The annihilation cross section is 
\be \langle \sigma v_{\mathrm{rel}}\rangle_{F_6F_6\rightarrow Z'Z'}\simeq \fr{g'^4_N}{16\pi m^2_6}\gg 1\ \mathrm{pb},\ee since $g'_N\simeq 1$ and that $m_6$ is radically below the weak scale. The $F_6$ density is negligible, \be \Om_{F_6} h^2\simeq \fr{0.1\ \mathrm{pb}}{\fr 1 2 \langle \sigma v_{\mathrm{rel}}\rangle_{F_6F_6\to Z'Z'}} < 2.58\times 10^{-6},\ee for $m_{F_6}<10$ GeV. Here the factor $\fr 1 2 $ associated with the annihilation cross-section arises because $F_6$ is a Dirac fermion and that the $F_6$ density refers to the sum of the abundances for $F_6$ and $\bar{F}_6$. Such factor and density definition also apply to every Dirac fermion candidate below, which should be understood.  

\begin{figure}[h]
\bc
\resizebox{0.5\textwidth}{!}{%
  \includegraphics{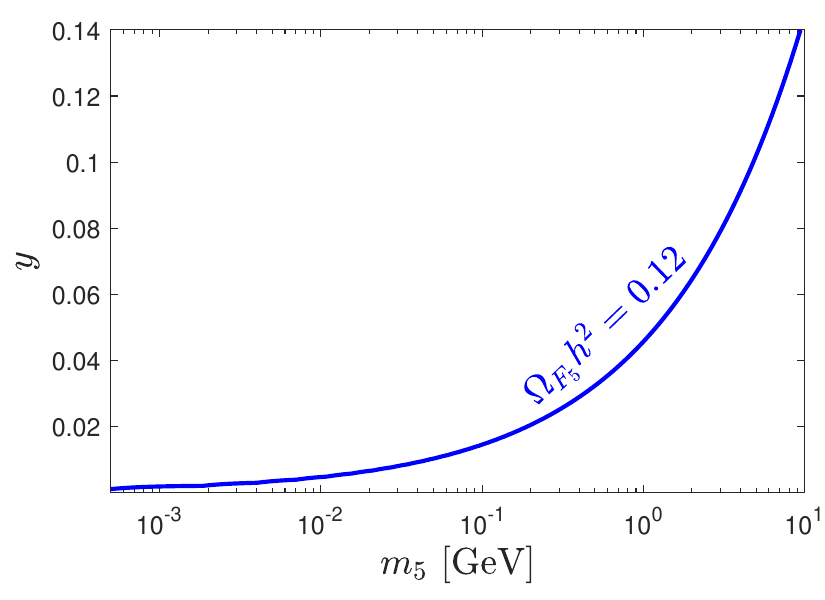}}
\caption{Dark matter self-coupling and mass contoured for the correct abundance, taking $m_7=2m_5$.} 
\label{OmF5h2}
\ec     
\end{figure}

Hence, the cold dark matter abundance is approximated by the $F_5$ contribution,    
\be \Omega_{\text{DM}}h^2 =\Omega_{F_5}h^2+\Omega_{F_6}h^2\simeq \Omega_{F_5}h^2.\ee Applying the Feynman rules for the left diagram in Fig.~\ref{darkcharge1}, we obtain the annihilation cross-section for $F_5$,
\be
\langle\sigma v_{\mathrm{rel}}\rangle_{F_{5}F_5\rightarrow F_6F_6}\simeq \frac{|y|^4(m_5+m_6)^2}{8\pi (m_7^2-m_6^2+m_5^2)^2}\left(1-\frac{m_6^2}{m_5^2}\right)^{1/2}.\label{anni56}
\ee
Using the approximation $m_6\approx m_5$, the data for dark matter density \cite{Zyla:2020zbs} \be \Omega_{\text{DM}}h^2\simeq \fr{0.1\ \mathrm{pb}}{\fr 1 2 \langle\sigma v_{\mathrm{rel}}\rangle_{F_5 F_5\to F_6 F_6}} \simeq 0.12\ee is recovered if $\langle\sigma v_{\mathrm{rel}}\rangle_{F_5 F_5\to F_6 F_6}\simeq 1.67$ pb,\footnote{Let us note that $1.67\ \mathrm{pb}\simeq 5\times 10^{-26}\ \mathrm{cm}^3\mathrm{s}^{-1}$ as the benchmark value used before.} or 
\be |y|\simeq 5.88\times 10^{-4}\left(\fr{m_7}{2m_5}\right)\left(\fr{m_5}{m_e}\right)^{1/2} \left(\fr{\delta m^2}{m^2_5}\right)^{-1/8}.\ee The dark matter self-coupling depends on $m_7/m_5$ and $m_5/m_e$, having a lower bound $|y|>10^{-3}$, for $m_7>2m_5$ and $m_5>m_e$, where note that $\delta m^2/m^2_5\equiv (m^2_5-m^2_6)/m^2_5\sim 1\%$. As an instance, we make a contour of the dark matter self-coupling $y$ and the $F_5$ mass $m_5$ in Fig. \ref{OmF5h2}.

Note that $F_5$ may annihilate to $Z'$ via 1-loop diagrams with both $\Phi_7$ and $F_6$ running in the loop. The amplitude is then evaluated as $M_{\mathrm{rad}}\sim \fr{y^2m_5}{m^2_{6,7}}\fr{g'^2_N}{16\pi^2}$, which is substantially smaller than the given tree-level amplitude, $M_{\mathrm{tree}}\sim \fr{y^2 m_{5,6}}{m^2_7}$, due to the loop factor suppression $\sim 1/16\pi^2$ and $g'_N\simeq 1$. The tree-level annihilation of $F_5$ to $F_6$ sets the relic density.  

\subsection{\label{wimps} Large $U(1)_N$ coupling: Implication for co-existed WIMPs}

WIMP is a natural solution of the thermal dark matter paradigm, which in the past it contributed to the thermal bath of the early universe, then decoupled, i.e. freeze-out, with a present-day density, when its annihilation rate into usual particles matches the Hubble expansion rate. Many standard model extensions imply existence of a single WIMP, such as supersymmetry and extradimension \cite{Jungman:1995df,Bertone:2004pz}, as well as the gauge approach \cite{Dong:2013ioa,Dong:2014esa,Dong:2014wsa,Dong:2015rka,Dong:2015dxw,Huong:2016ybt,Huong:2018ytz,Huong:2019vej,Dinh:2019jdg,Dong:2016sat,Dong:2017ayu,VanDong:2020nwb}. However, this work adds that the dequantization of electric charge unravels a picture of structured WIMPs. It is noted that the versions of dark matter under consideration actually contain two simultaneously stabilized WIMPs. Hence, it is sufficiently to investigate one of them, say the model with $\delta=1$, with the candidates in Table \ref{canZ6}, which have the minimal basic hyperdark charges and a large $U(1)_N$ gauge coupling. We also assume the $F_{5,6}$ masses to be beyond the weak scale. 

Dark matter pair annihilation in the present case includes those given similarly to Fig. \ref{darkcharge1}, plus the extra $s$-channel diagrams (not depicted) for $F_6$ annihilation to the standard model particles (leptons, quarks, Higgs and gauge bosons) exchanged by $Z'$, namely $F_6F^c_6\to f f^c,ZH$ where $f$ indicates to the usual leptons and quarks. Notice that $F_5$ does not annihilate to the usual particles, since its dark charge vanishes. 

It is straightforward to determine the annihilation cross section of $F_6$, such as 
\bea
\langle\sigma v_{\mathrm{rel}}\rangle_{F_6F_6\to \mathrm{all}}&\simeq& \frac{\theta(m_6-m_{Z^\prime}) g_N^4 m_6^2}{324\pi(m^2_{Z^\prime}-2m^2_6)^2} \left(1-\frac{m^2_{Z^\prime}}{m^2_6}\right)^{3/2}\crn
&&+\frac{g_N^2 m_6^2\sum_f N_C(f)\{[g_V^{Z^\prime}(f)]^2+[g_A^{Z^\prime}(f)]^2\} }{9\pi(m_{Z^\prime}^2-4m_6^2)^2}\crn
&&+\frac{g_N^2 g_{Z'ZH}^2m^2_6}{144\pi(m_{Z^\prime}^2-4m_6^2)^2m^2_Z},\label{adtnt1}
\eea
where $g^{Z'}_{V,A}(f)$ were supplied in Table \ref{Zp}, while the $HZZ'$ coupling is given by $g_{Z'ZH}\simeq -g_Nv/2$.

Since the dark matter components $F_{5,6}$ now have arbitrary masses above the weak scale, the conversion between $F_{5,6}$ is either $F_5F_5^c\to F_6F_6^c$ if $m_5>m_6$ or $F_6F_6^c\to F_5F_5^c$ if $m_5<m_6$. The former process is described by the left diagram in Fig. \ref{darkcharge1}, in which the result in Eq. (\ref{anni56}) properly applies for this case. The latter, or inverse process, is given by a $t$-channel diagram, similar to the previous process by replacement $F_5\leftrightarrow F_6$ with the same $\Phi_7$ mediator, leading to an annihilation cross section, 
\be
\langle\sigma v_{\mathrm{rel}}\rangle_{F_{6}F_6\to F_5 F_5}\simeq \frac{|y|^4(m_5+m_6)^2}{8\pi (m_7^2-m_5^2+m_6^2)^2}\left(1-\frac{m_5^2}{m_6^2}\right)^{1/2}. \label{anni65}
\ee 

It is noted that for the case $m_6>m_5$, the annihilation cross sections in Eqs. (\ref{adtnt1}) and (\ref{anni65}) set the $F_6$ density. However, $F_5$ never annihilates after the temperature of the universe falls below the $F_6$ mass. Thus, it has a large density, overpopulating the universe. This case should be discarded. Hence, we consider only $m_5>m_6$. And, the $F_5$ density is governed by its annihilation to $F_6$, i.e. \be \Om_{F_5} h^2\simeq \fr{0.1\ \mathrm{pb}}{\fr 1 2 \langle \sigma v_{\mathrm{rel}}\rangle_{F_5F_5\to F_6 F_6}},\ee with the annihilation cross-section given by Eq. (\ref{anni56}), while the $F_6$ density is determined by its annihilation to the $Z'$ and/or standard model particles, \be \Om_{F_6} h^2\simeq \fr{0.1\ \mathrm{pb}}{\fr 1 2 \langle \sigma v_{\mathrm{rel}}\rangle_{F_6F_6\to \mathrm{all}}},\ee with the annihilation cross-section obtained by Eq. (\ref{adtnt1}). The total dark matter density is $\Om_{\mathrm{DM}} h^2=\Om_{F_5} h^2+\Om_{F_6} h^2$, where both the components contribute, unlike the above model for XENON1T.       

\begin{figure}[h]
\resizebox{0.5\textwidth}{!}{%
\includegraphics{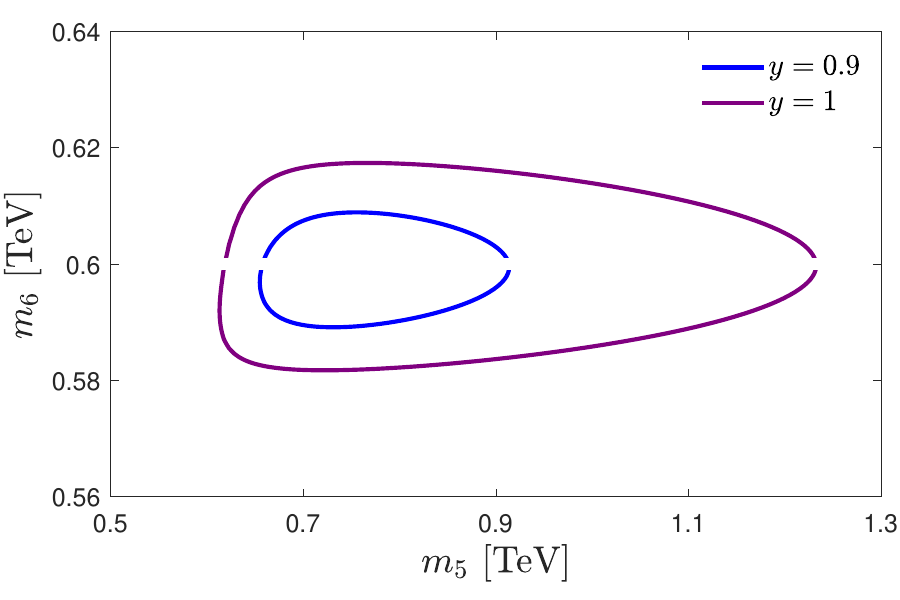}}

\vs

\resizebox{0.5\textwidth}{!}{%
\includegraphics{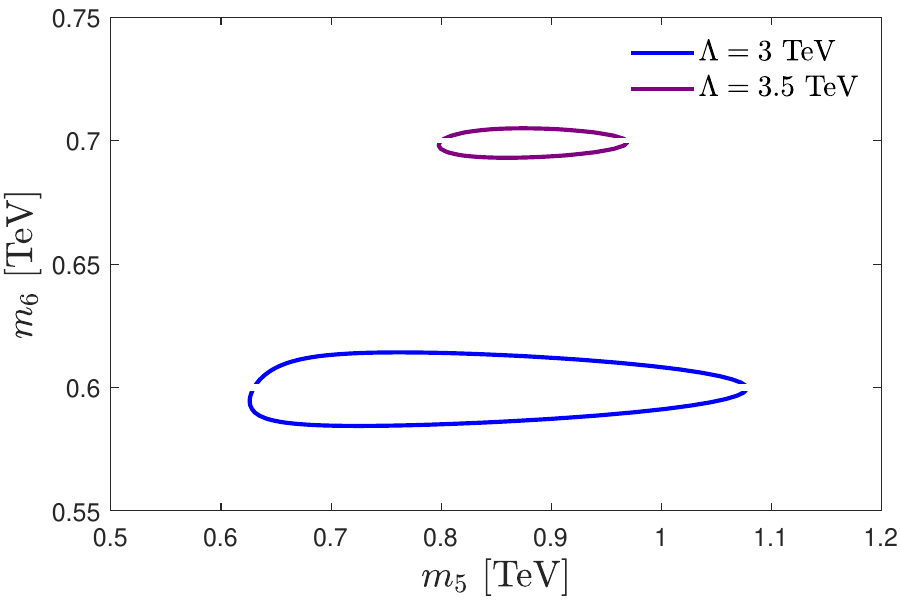}}
\caption[]{\label{figfdt} Total dark matter density contoured as functions of component dark matter masses for different choices of $y,\La$; the value of other parameters relevant to each panel is detailedly supplied in the text.}
\end{figure}

The parameters that significantly govern the dark matter observables are the dark field self-coupling $y$, the $U(1)_N$ gauge coupling $g_N$, and the new physics scale $\La$, besides the dark matter masses $m_{5,6}$ and the mediator mass $m_7$. Let us make contours of $\Om_{\mathrm{DM}}h^2=0.12$ inspired by the experiment \cite{Zyla:2020zbs} as functions of $m_{5,6}$ according to several choices of the remaining parameters. Namely, in Fig. \ref{figfdt} the upper panel is plotted for $x \equiv m_7/(m_5+m_6) = 1.1$, $g_N=0.2$, $\Lambda=3$ TeV, and $y=0.9$ and 1 corresponding to each curve as marked in the panel. Whereas, also in Fig. \ref{figfdt}, the lower panel is plotted for $y=0.95$, $x=1.1$, $g_N=0.2$, and the curves according to the values of $\La=3$ and $3.5$ TeV, respectively.    
 Note that the disconnected regions on each curve are due to a $Z'$ resonance in the density, at which $m_6=\fr 1 2 m_{Z'}$, which reduces the density to zero, thus suppressed for correct abundance. 

Since $F_5\sim (1,1,0,0)$ is sterile, it does not interact with detectors in direct detection experiment. Effect of dark matter in the direct detection comes only from a potential scatter of the $F_6$ component with nuclei in a large detector.\footnote{Since this WIMP is non-relativistic, yielding a small electronic recoil energy, it cannot explain the XENON1T excess.} Indeed, the effective Lagrangian describing the $F_6$-quark interaction can be derived from the $t$-channel exchange diagram by the new gauge boson $Z'$ to be 
\be
\mathcal{L}_{\text{eff}}\supset \bar{F}_6 \gamma^\mu F_6 \left[\bar{q}\gamma_\mu (\alpha_q P_L+\beta_q P_R)q\right],
\ee
where $P_{L,R}=(1\mp\gamma_5)/2$, and \be \alpha_{u,d}=-\beta_u/2=\beta_d/4=-g_N^2/18m_{Z'}^2.\ee Hence, we obtain the spin-independent (SI) cross-section for the scattering of $F_6$ on a nucleon as
\be
\sigma^{\text{SI}}_{F_6-\text{nucleon}}=\frac{4m^2_{\text{nucleon}}}{\pi A^2}[\lambda_p Z+\lambda_n (A-Z)]^2,
\ee
where \bea &&\lambda_p=[2(\alpha_u+\beta_u)+\alpha_d+\beta_d]/8,\crn 
&&\lambda_n=[\alpha_u+\beta_u+2(\alpha_d+\beta_d)]/8.\eea Above, $Z$ is the nucleus charge and $A$ is the total number of nucleons in the nucleus. Because the dark matter components partially contribute to the total density, the effective SI cross-section of $F_6$ is given by
\be \sigma^{\text{SI}}_{\text{eff}}(F_6)= \frac{\Omega_{F_6}h^2}{\Omega_{\text{DM}}h^2} \sigma^{\text{SI}}_{F_6-\text{nucleon}}. \ee
\begin{figure}[h]
\bc
\resizebox{0.6\textwidth}{!}{%
  \includegraphics{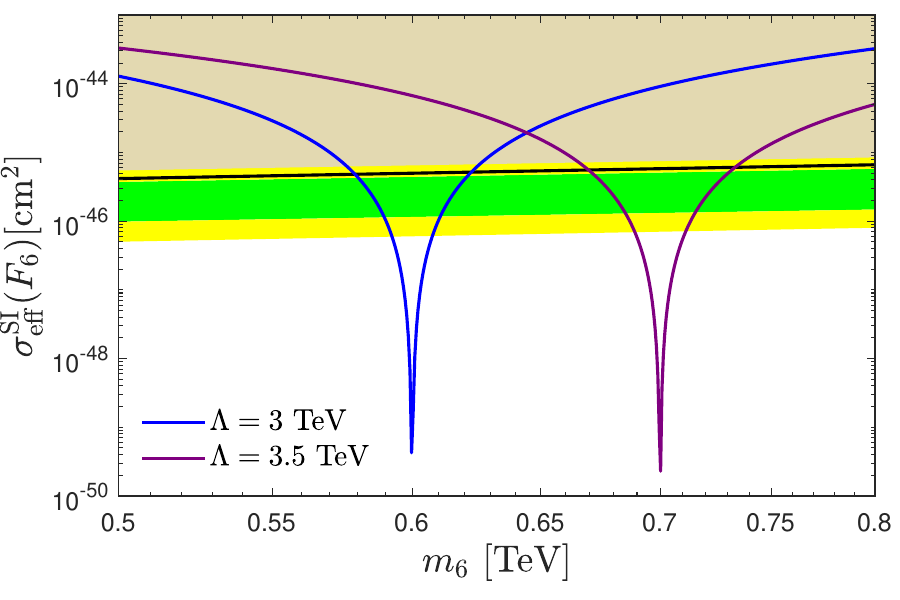}}
\caption{The SI $F_6$-nucleon effective scattering cross-section limit as a function of $F_6$ mass according to the several choices of $\Lambda$, where the excluded region (light brown) lies above the experimental (green and yellow) bands.}
\label{SIF6}      
\ec
\end{figure}

Taking $A=131$ and $Z=54$ according to the Xe nucleus and $m_{\text{nucleon}}\simeq 1$ GeV, in Fig. \ref{SIF6} we plot the SI cross-section of $F_6$ corresponding to the previous choices of $\Lambda=3$ and 3.5 TeV, $y=0.95$, $x=1.1$, and $g_N=0.2$, which resulted in Fig. \ref{figfdt} lower panel. We always assume that the two dark matter components give the correct relic density. The XENON1T experimental bounds \cite{Aprile:2017iyp,Aprile:2018dbl} have also been included to the plot. Combining the results in Fig. \ref{figfdt} lower panel and Fig. \ref{SIF6}, the viable dark matter mass regime is either $0.63 \text{ TeV}<m_5<1.07 \text{ TeV}$ and $0.58 \text{ TeV}<m_6<0.61 \text{ TeV}$ for $\Lambda=3$ TeV, or $0.80 \text{ TeV}<m_5<0.97 \text{ TeV}$ and $0.69 \text{ TeV}<m_6<0.71 \text{ TeV}$ for $\Lambda=3.5$ TeV, but always ensuring that $m_5>m_6$.    

\section{\label{conl}Conclusion}

The dequantization of electric charge leads to the existence of a noncommutative dark charge and that the full gauge symmetry of the model takes the form, $SU(3)_C\otimes SU(2)_L\otimes U(1)_Y \otimes U(1)_N$. This setup not only yields suitable neutrino masses, but also implies the novel schemes of multi-component dark matter. The dark matter stability is ensured by the residual dark charge identical to an even $Z_k$ symmetry, a remnant of the gauge symmetry after symmetry breaking. Additionally, the dark matter components can have an arbitrary mass, despite the fact that they and the normal particles transform nontrivially under the discrete symmetry, that cannot decay as a consequence of the color and electric charge conservation, combined with $Z_k$. With this mechanism, we have pointed out the simplest models for two-component dark matter. Additionally, we have proved that they can address the XENON1T anomaly recently observed, as the second dark matter component is boosted in annihilation of the first dark matter component. Alternatively, these models can contain two WIMPs with relevant masses above the weak scale, satisfying the relic density and direct detection bounds.  
\section*{Acknowledgments}

We would like to thank Dr. Cao H. Nam (Phenikaa University) for the useful discussions. This research is funded by Vietnam National Foundation for Science and Technology Development (NAFOSTED) under grant number 103.01-2019.353.

\bibliographystyle{JHEP}

\bibliography{combine}

\end{document}